\documentclass[aps,prd,twocolumn,nofootinbib,twoside,floatfix]{revtex4}

\usepackage{amsmath}
\usepackage{amssymb}
\usepackage{graphicx}
\usepackage{subfig}
\usepackage{fancyhdr}
\usepackage{caption}

\newcommand{\Cal}[1]{\ensuremath{\mathcal{#1}}}

\newcommand{\eqn}[1]{Eqn. \eqref{#1}}
\newcommand{\eqns}[1]{Eqns. \eqref{#1}}
\newcommand{\Cite}[1]{Ref. \cite{#1}}
\newcommand{\Cites}[1]{Refs. \cite{#1}}
\newcommand{\Sec}[1]{Sec. \ref{#1}}
\newcommand{\fig}[1]{Fig. \ref{#1}}
\newcommand{\figs}[1]{Figs. \ref{#1}}
\newcommand{\ph}[1]{\phantom{#1}}

\newcommand{\G}{\ensuremath{\Gamma}}
\newcommand{\GF}{\ensuremath{\Gamma_F}}
\newcommand{\dG}{\ensuremath{\delta\Gamma}}
\newcommand{\dg}{\ensuremath{\delta\Gamma^{(1)}}}
\newcommand{\Gb}{\ensuremath{\widetilde{\Gamma}}}
\newcommand{\dGb}{\ensuremath{\widetilde{\delta\Gamma}}}
\newcommand{\dgb}{\ensuremath{\widetilde{\delta\Gamma}^{(1)}}}

\newcommand{\W}{\ensuremath{W}}
\newcommand{\Wi}{\ensuremath{W^{-1}}}
\newcommand{\dW}{\ensuremath{\delta W}}
\newcommand{\dw}{\ensuremath{\delta W^{(1)}}}

\newcommand{\p}{\ensuremath{\partial}}
\newcommand{\pp}{\ensuremath{\partial^\prime}}

\newcommand{\avg}[1]{\ensuremath{\langle ~#1~ \rangle}}
\newcommand{\eavg}[1]{\ensuremath{[ ~#1~ ]}_{ens}}

\newcommand{\be}{\begin{equation}}
\newcommand{\ee}{\end{equation}}

\newcommand{\ti}[1]{{\tilde #1}}
\newcommand{\wti}[1]{{\widetilde #1}}
\newcommand{\vphi}{\varphi}
\newcommand{\om}{\omega}
\newcommand{\com}{{\hat\omega}}
\newcommand{\cchi}{{\hat\chi}}
\newcommand{\cdelta}{{\hat\delta}}
\newcommand{\ctheta}{{\hat\Theta}}
\newcommand{\cv}{{\hat V}}

\newcommand{\Wxx}[2]{\ensuremath{\Cal{W}{}^{#1^\prime}_{#2}(x^\prime,x)}}
\newcommand{\Wxxinv}[2]{\ensuremath{\Cal{W}{}^{#1}_{#2^\prime}(x,x^\prime)}}

\newcommand{\Mbar}{\ensuremath{\bar{\Cal{M}}}}

\begin{document}
\title{Backreaction of Cosmological Perturbations in Covariant
  Macroscopic Gravity}
\author{Aseem Paranjape}
\email{aseem@tifr.res.in}
\affiliation{Tata Institute of Fundamental Research, Homi Bhabha
Road, Colaba, Mumbai - 400 005, India\\}
\begin{abstract}
\noindent
The problem of corrections to Einstein's equations arising from
averaging of inhomogeneities (``backreaction'') in the cosmological
context, has gained considerable attention recently. We present results
of analysing cosmological perturbation theory in the framework of
Zalaletdinov's fully covariant Macroscopic Gravity. We show that this
framework can be adapted to the setting of cosmological perturbations
in a manner which is free from gauge related ambiguities. We derive
expressions for the backreaction which can be readily applied in
\emph{any} situation (not necessarily restricted to the linear
perturbations considered here) where the \emph{metric} can be
brought to the perturbed FLRW form. In particular these expressions
can be employed in toy models studying nonlinear structure formation,
and possibly also in $N$-body simulations. Additionally, we present 
results of example calculations which show that the backreaction
remains negligible well into the matter dominated era.
\end{abstract}

\maketitle
\section{Introduction}
\label{sec:intro}
\noindent
It is known \cite{ellis} that in order to apply the equations of
Einstein's General Relativity at the large length scales of interest
in cosmology, it is necessary to first perform a smoothing or
averaging operation, which will generate nontrivial corrections in
these nonlinear equations. There is a debate in the literature
concerning two basic questions regarding this smoothing operation :
(a) How does one obtain consistent and observationally relevant
variables associated with the corrections due to averaging or
``backreaction'', and (b) what is the magnitude of this backreaction? 

This discussion has had a long history \cite{avg,siegel,avg-bran},
although in recent times attention has been focussed on two promising 
candidates for a consistent \emph{nonperturbative} averaging
procedure, namely the spatial averaging of scalars due to Buchert
\cite{buchert} and the covariant approach due to Zalaletdinov
\cite{zala1,zala2}. Perhaps owing to its appealing simplicity of
implementation, the approach due to Buchert has attracted a
significant amount of attention both in the context of cosmological
perturbation theory \cite{pert-buchertavg,kolb,behrend} and with
regards fully nonlinear calculations
\cite{avg-buchert,rasanen1,rasanen2}, although very interesting
results have been derived using Zalaletdinov's covariant procedure as
well \cite{coley,vdhoogen}. The most fascinating aspect of these
studies has been the possibility that the phenomenon of Dark Energy
\cite{kolb,rasanen2} and also Dark Matter \cite{vdhoogen} might be
attributed to the backreaction from averaging (see \Cite{buchert-rev}
for a recent review of the subject).  

The physical relevance of the averaged variables defined in Buchert's
formalism, has been questioned in the literature \cite{wald}. 
It has been argued \cite{wald,smalleffect} that
effects of inhomogeneities which only perturbatively affect the (on
average homogeneous) metric of the universe, \emph{cannot} lead to
effects large enough to account for the inferred acceleration of the
universe from, say supernovae observations \cite{SNe}. This has been
countered by the argument \cite{rasanen2} that perturbation theory may
break down during the epoch of fully nonlinear structure
formation. Recently it was shown \cite{SphColl} in the context of a
specific model of spherical collapse, that an explicit coordinate
transformation could be found, which brought the metric to the
perturbed Friedmann-Lema\^itre-Robertson-Walker (FLRW) form, 
\be
ds^2 = -(1+2\vphi)d\tau^2 + a^2(\tau)(1-2\psi)d\vec{x}^2\,,
\label{eq1}
\ee
satisfying all conditions required for a perturbation formalism in
$\vphi$ and $\psi$ to hold, \emph{even} in the regime of fully
nonlinear collapse.

In such a context, it is important to have a consistent formalism,
free from issues such as gauge artifacts, in which one has derived
expressions for the backreaction which can be applied in a
straightforward manner to any model in which the metric can be brought
to the form \eqref{eq1}. Since the Buchert framework by construction
is best adapted to coordinates comoving with the matter, most
applications using perturbation theory in this framework have focussed
on the synchronous and comoving gauge, although recently Behrend et
al. \cite{behrend} have also performed calculations in the conformal
Newtonian gauge. However, as pointed out elsewhere
\cite{SphColl}, calculations using the Buchert framework in
perturbation theory necessarily face the ambiguity of dealing with
\emph{two} scale factors : one is the scale factor with which the
perturbed FLRW metric is defined, and the other is the volume averaged
scale factor defined by Buchert \cite{buchert}, and it is not clear
which of these scale factors is the observationally relevant one.

In this paper we shall deal with the covariant framework of
Zalaletdinov's Macroscopic Gravity (MG), and its spatial averaging
limit proposed in \cite{spatavglim}. We will see that here, one has a
well defined averaged metric with a corresponding \emph{uniquely
  defined} scale factor. Further, we will argue that the structure of
MG allows us to completely specify a consistent, observationally
relevant averaging operation adapted for perturbations to the FLRW
geometry. The final expressions obtained for the backreaction can be
applied directly (or with some straightforward modifications), to
\emph{any} situation in which the metric can be brought to the form
\eqref{eq1} or a suitable generalisation thereof.  

The organisation of this paper is as follows : \Sec{sec:MG} presents a
brief recap of the basic ideas underlying MG and some
useful expressions from standard cosmological perturbation theory
(PT), together with a proposal for practically estimating the effect
of the backreaction on the time evolution of the FLRW scale
factor. Sec. \ref{sec:avgVP} presents details of the MG averaging
procedure adapted to cosmological PT, including general expressions
for the leading order backreaction terms, with a discussion of gauge
related issues and the definition of the averaging 
operator. The heart of the paper is in \Sec{sec:corrscal}, where we
derive final expressions for the backreaction, both in real space and
Fourier space, which can be directly utilised in model
calculations. While these expressions use a few simplifying
restrictions, these can be lifted if necessary in a completely
straightforward manner. \Sec{sec:exs} contains example calculations in
first order PT, which show that the magnitude of the backreaction is,
as expected, negligible compared to the homogeneous energy density of
matter in the radiation dominated era and for a significant part of
the matter dominated era. We conclude in \Sec{sec:discuss} with some
final comments. Throughout the paper, lower case Latin indices
$a,b,c$... will refer to spacetime indices $0,1,2,3$, and upper case
Latin indices $A,B,C,$... to spatial indices $1,2,3$. The speed of
light $c$ is set to unity, and a prime refers to a derivative with
respect to conformal time unless stated otherwise.

\section{Covariant Macroscopic Gravity (MG) and Cosmology}   
\label{sec:MG}
\subsection{Recap of MG formalism}
\label{sec:MG:sub:recap}
\noindent
In this section we present a rapid overview of the generally
covariant averaging formalism of Macroscopic Gravity (MG), developed
by Zalaletdinov and coworkers \cite{zala1,zala2,mars,zala-pert}, and
its spatial averaging limit proposed in \cite{spatavglim}. The reader
is referred to these papers for more details.

The MG formalism uses a bilocal operator \Wxx{a}{j}, called the
coordination bivector, to define a covariant spacetime averaging
operation on tensors in a spacetime manifold \Cal{M}. [The prime here
  is being used to distinguish two different spacetime points, and
  must not be confused with a derivative with respect to conformal
  time. ] The various properties that this operator must satisfy, and
a proof of its existence can be found in \Cites{zala1,zala2,mars}. The
form of the coordination bivector used in all MG calculations is
\be
\Wxx{a}{b} = \left.\frac{\p x^a}{\p x^j_{\rm VP}}\right|_{x^\prime}
\left.\frac{\p x^j_{\rm VP}}{\p x^b}\right|_x \,,
\label{MG-W}
\ee
where $x^j_{\rm VP}$ refers to a coordinate system in which the metric
determinant is constant. The class of such coordinate systems is
called volume preserving (VP). Not surprisingly, the entire formalism
of MG simplifies considerably in VP coordinate systems, in which the
coordination bivector reduces to the Kronecker delta 
\be
\Wxx{a}{j}|_{VPC} = \delta{}^a_j\,.
\label{MG-W-VP}
\ee
We will see later that VPC systems play an important role in
consistently setting up cosmological perturbation theory in the
context of MG.

The average of a tensor $P{}^a_b$ over a finite spacetime domain
$\mathbf{\Sigma}$ is given by
\begin{align}
\bar P{}^a_b(x) &= \avg{\wti{P}{}^a_b}(x) =
\frac{1}{V_{\mathbf{\Sigma}}}   
\int_{\mathbf{\Sigma}}{d^4x^\prime\sqrt{-g^\prime}
  \ti{P}{}^a_b(x^\prime,x) }~~;\nonumber\\
 V_{\mathbf{\Sigma}} &=
\int_{\mathbf{\Sigma}}{d^4x^\prime\sqrt{-g^\prime}}\,,   
\label{MG1}
\end{align}
where $\ti{P}{}^a_b(x^\prime,x)$ is the bilocal extension of $P{}^a_b$
defined as
\be
\ti{P}{}^a_b(x^\prime,x) = \Wxxinv{a}{i}
P{}^{i^\prime}_{j^\prime}(x^\prime) \Wxx{j}{b}\,.
\label{MG2}
\ee
The averaging operation when appropriately applied to the connection
$\G^a_{bc}$ on \Cal{M}, gives an averaged connection
$\bar\Gamma^a_{bc}$ which is taken to be the connection on an averaged
manifold \Mbar. In other words, the connection on \Mbar\ satisfies the
condition
\be
\avg{\Gb^a_{bc}} = \bar\G^a_{bc}\,.
\label{MG-avgcond}
\ee
The metric $G_{ab}$ associated with the averaged
connection can be assumed to be the average  of the inhomogeneous
metric $g_{ab}$ on \Cal{M}, i.e. $G_{ab}=\bar g_{ab}$. (We will see
later that in the perturbative setting this amounts to a very natural
condition on the perturbations.) Averaging the Einstein equations on
\Cal{M}\ leads to the equations satisfied by the averaged metric,
which can be written as       
\be
E{}^a_b = 8\pi G_N\ T{}^a_b + {}^{(grav)}T{}^a_b\,.
\label{MG3}
\ee
Here $E{}^a_b$ is the Einstein tensor constructed from the metric
$G_{ab}$ and its inverse $G^{ab}$, $T{}^a_b$ is the averaged
energy-momentum tensor, $G_N$ is Newton's gravitational constant, and
${}^{(grav)}T{}^a_b$ is a tensorial correlation object which acts like
an effective gravitational energy-momentum tensor. For brevity we shall
omit its detailed definition, referring the reader to
\Cite{zala1,zala2}, and only note that its broad structure can be
symbolically represented as   
\be
{}^{(grav)}T\sim \avg{\Gb^2} - \avg{\Gb}^2\,,
\label{MG4}
\ee
where $\Gb$ symbolically denotes the bilocal extension of the  
Christoffel connection on \Cal{M}. In general the total
energy-momentum tensor $(8\pi G_N T{}^a_b + {}^{(grav)}T{}^a_b)$ is
covariantly conserved,
\be
(8\pi G_N T{}^a_b + {}^{(grav)}T{}^a_b)_{;a} = 0\,,
\label{MG-conserv}
\ee 
with the semicolon denoting the covariant derivative with respect to
the averaged geometry ($\bar\Gamma^a_{bc}$ and $G_{ab}$).

In \Cite{spatavglim} it was argued that in the cosmological context,
it is essential to consider a spatial averaging limit of the covariant
averaging used in MG. The simplest way to see this is to note that the
homogeneous and isotropic FLRW spacetime \emph{must} be left invariant
under the averaging operation, and this is only possible if the
averaging is tuned to the uniquely defined spatial slices of constant
curvature in the FLRW spacetime. A related statement also applies to
perturbation theory in general. The application of the MG formalism
to perturbation theory has been discussed in \Cites{zala2,zala-pert},
which also compare the results of MG in the perturbative scenario,
with those obtained by Isaacson \cite{isaacson} in the study of the
short wavelength limit of gravitational waves. A crucial observation
\cite{zala-pert} is that what one normally refers to as the
``background'' in perturbation theory, must be defined by an averaging
procedure. In particular, the background must remain invariant under
the averaging operation. In \Cite{spatavglim} no \emph{a priori}
assumption was made about the \emph{form} of the inhomogeneous metric,
and hence certain assumptions had to be made about the choice of
spatial slicing used to define the spatial averaging. In the case at
hand, namely when the inhomogeneous metric is taken to be a
perturbation around the FLRW metric, we will see that the situation
simplifies to some extent and a consistent spatial averaging operation
can be identified.  

Throughout this paper we will assume that the metric of the universe is
a perturbation around the FLRW metric given by
\be
ds^2 = a^2(\eta)\left( -d\eta^2 + \gamma_{AB}dx^Adx^B \right)\,.
\label{FLRW-metric}
\ee
Here $a$ is the scale factor and $\eta$ is the conformal time
coordinate related to cosmic time $\tau$ by the differential relation 
\be
d\tau = a(\eta)d\eta
\label{conf-time}
\ee
In \eqn{FLRW-metric} we have allowed the spatial metric to have the
general form $a^2\gamma_{AB}$ where $\gamma_{AB}$ is the metric of a
$3$-space of constant curvature. For the calculations in this paper we
will assume a flat FLRW background in coordinates such that
$\gamma_{AB}=\delta_{AB}$; however for future reference we shall
present certain expressions in terms of the more general spatial
metric.  

Assuming the averaged metric to have the form \eqref{FLRW-metric},
and the averaged energy-momentum tensor to have the form
\be
T{}^a_b = (\rho+p)\bar v^a\bar v_b + p\delta{}^a_b
\,, 
\label{MG6}
\ee
where $\bar v^a$ is the timelike 4-vector which defines the
homogeneous spatial slices of the FLRW spacetime, and $\rho(\tau)$ and 
$p(\tau)$ are respectively the homogeneous energy density and pressure
corresponding to the averaged matter distribution, the modified
cosmological equations obtained by averaging a general (i.e. not
necessarily perturbed FLRW) spacetime, can be shown to reduce to the
following (see Eqns. 87 of \Cite{spatavglim}) 
\begin{subequations}
\begin{align}
&\left(\frac{1}{a}\frac{da}{d\tau}\right)^2 = \frac{8\pi G_N}{3}\rho -
\frac{1}{6}\left[ \Cal{P}^{(1)} + \Cal{S}^{(1)} \right]  \,,
\label{MG7a}\\ &\nonumber\\
&\frac{1}{a}\frac{d^2a}{d\tau^2} = -\frac{4\pi
  G_N}{3}\left(\rho+3p\right) + \frac{1}{3}\left[
  \Cal{P}^{(1)} + \Cal{P}^{(2)} +
  \Cal{S}^{(2)} \right] \,,
\label{MG7b}
\end{align}
\label{MG7}
\end{subequations}
where the combinations $(\Cal{P}^{(1)} + \Cal{S}^{(1)})$ and $(
\Cal{P}^{(1)} + \Cal{P}^{(2)} +  \Cal{S}^{(2)})$ are generally
covariant scalars defined by the relations (see Eqns. 88 of
\Cite{spatavglim}, with $f=a$)  
\begin{subequations}
\begin{align}
&\Cal{P}^{(1)} = \frac{1}{a^2}\left[
  \avg{\wti{\Gamma}{}^A_{0A}\wti{\Gamma}{}^B_{0B}} -
  \avg{\wti{\Gamma}{}^A_{0B}\wti{\Gamma}{}^B_{0A}} - 6\Cal{H}^2\right]
  \,, 
\label{MG8a} \\&\nonumber\\
&\Cal{S}^{(1)} =
\avg{\wti{g}^{JK}}\left[\avg{\wti{\Gamma}{}^A_{JB}\wti{\Gamma}{}^B_{KA}}
  - \avg{\wti{\Gamma}{}^A_{JA}\wti{\Gamma}{}^B_{KB}}\right] \,,
\label{MG8b} \\&\nonumber\\
&\Cal{P}^{(2)} + \Cal{P}^{(1)} =
-\frac{1}{a^2}\avg{\wti{\Gamma}{}^A_{0A}\wti{\Gamma}{}^0_{00}} -
\avg{\wti{g}^{JK}}\avg{\wti{\Gamma}{}^0_{JA}\wti{\Gamma}{}^A_{0K}}
\nonumber\\ 
&~~~~~~~~~~~~~~~~~~+ \frac{6\Cal{H}^2}{a^2} \,, 
\label{MG8c} \\&\nonumber\\
&\Cal{S}^{(2)} =
\frac{1}{a^2}\avg{\wti{\Gamma}{}^A_{00}\wti{\Gamma}{}^0_{A0}} +
\avg{\wti{g}{}^{JK}}\avg{\wti{\Gamma}{}^0_{J0}\wti{\Gamma}{}^A_{KA}}
\,, \label{MG8d}
\end{align}
\label{MG8}
\end{subequations}
where we have defined 
\be
\Cal{H} = \frac{1}{a}\frac{da}{d\eta} \equiv \frac{a^\prime}{a} \,, 
\label{FLRW-H-1}
\ee
and accounted for the fact that in general, the average of the inverse
inhomogeneous metric, need not equal the inverse metric
$G^{ab}$. However, we will soon see that in the perturbative setting
we can in fact set these two tensors to be equal via a very natural
condition on the perturbations. The index $0$ in \eqns{MG8} refers to
the conformal time $\eta$. The averaging in \eqns{MG8} is assumed
to be a spatial averaging in an unspecified spatial slicing in the
inhomogeneous manifold \Cal{M}; in \Sec{sec:avgVP} we will specify the
averaging procedure more exactly.

In addition, the following ``cross-correlation'' constraints must also
be satisfied by the inhomogeneities (see Eqns. 89 of \Cite{spatavglim}) 
\begin{subequations}
\begin{align}
&\frac{1}{a^2}\left[\avg{\wti{\Gamma}{}^0_{0A}\wti{\Gamma}{}^B_{B0}}
- 
  \avg{\wti{\Gamma}{}^0_{0B}\wti{\Gamma}{}^B_{A0}}\right] \nonumber\\
&\ph{\frac{1}{a^2}[]}
+  \avg{\wti{g}^{JK}}\left[
  \avg{\wti{\Gamma}{}^0_{JB}\wti{\Gamma}{}^B_{AK}}
  - \avg{\wti{\Gamma}{}^0_{JA}\wti{\Gamma}{}^B_{BK}} \right] = 0\,
\label{MG9a} \\&\nonumber\\
&\frac{1}{a^2}\left[\avg{\wti{\Gamma}{}^A_{00}\wti{\Gamma}{}^B_{B0}}
-
  \avg{\wti{\Gamma}{}^B_{00}\wti{\Gamma}{}^A_{B0}}\right] \nonumber\\
&\ph{\frac{1}{a^2}[]}
  +  \avg{\wti{g}^{JK}}\left[
  \avg{\wti{\Gamma}{}^A_{JB}\wti{\Gamma}{}^B_{0K}}
  - \avg{\wti{\Gamma}{}^A_{J0}\wti{\Gamma}{}^B_{BK}} \right] = 0\,
\label{MG9b}\\&\nonumber\\
&\frac{1}{a^2}\left[\avg{\wti{\Gamma}{}^A_{B0}\wti{\Gamma}{}^m_{0m}}
-
 \avg{\wti{\Gamma}{}^A_{m0}\wti{\Gamma}{}^m_{0B}} \right] \nonumber\\
&~~~+ \avg{\wti{g}^{JK}}
 \left[\avg{\wti{\Gamma}{}^A_{Jm}\wti{\Gamma}{}^m_{KB}} -
 \avg{\wti{\Gamma}{}^A_{JB}\wti{\Gamma}{}^m_{Km}} \right]
 \nonumber\\
&~~~=
 \delta{}^A_B\left[-\frac{1}{3}\left(\Cal{P}^{(2)} + \Cal{S}^{(2)} - 
   \Cal{S}^{(1)}\right) + \frac{4\Cal{H}^2}{a^2} \right] \,,
\label{MG9c}
\end{align}
\label{MG9}
\end{subequations}
where the lower case index $m$ in the last equation runs over all
spacetime indices $0,1,2,3$.

\subsection{Cosmological Perturbations and Gauge Transformations}
\label{app:pertavg}
\noindent
For ready reference, in this subsection we present expressions for the
metric, its inverse, and the Christoffel connection in \emph{first
  order} cosmological PT, in an arbitrary, unfixed gauge. The notation
we use is similar to that used in \Cite{bmr07}. We will also give
expressions for the first order gauge transformations of the
perturbation functions (see e.g. \Cite{bruni}). 

The first order perturbed FLRW metric in an arbitrary gauge can be
written as 
\begin{align}
&ds^2 = a^2(\eta)\left[ -(1+2\vphi)d\eta^2 + 2\om_Adx^Ad\eta +
  \right. \nonumber\\
&\ph{ds^2=a^2[-]}\left. \left((1-2\psi)\gamma_{AB} +
  \chi_{AB}\right)dx^Adx^B \right]\,.
\label{app1}
\end{align}
The functions $\vphi$ and $\psi$ are scalars under spatial coordinate 
transformations. The functions $\om_A$ and $\chi_{AB}$ can be
decomposed as follows
\begin{align}
\om_A = \p_A\om + \com_A ~~;~~
\chi_{AB} = D_{AB}\chi+2\nabla_{(A}\cchi_{B)} + \cchi_{AB}\,,
\label{app2}
\end{align}
where the parentheses indicate symmetrization; $D_{AB}$ is the
tracefree second derivative defined by 
\be
D_{AB} \equiv \nabla_A\nabla_B - (1/3)\gamma_{AB}\nabla^2 ~~;~~
\nabla^2 \equiv \gamma^{AB}\nabla_A\nabla_B\,,
\label{app3}
\ee
with $\nabla_A$ the covariant spatial derivative compatible with
$\gamma_{AB}$; and $\com_A$, $\cchi_A$ and $\cchi_{AB}$ satisfy
\be
\nabla_A\com^A = 0 = \nabla_A\cchi^A ~~;~~ \nabla_A\cchi{}^A_B = 0 = 
\cchi{}^A_A\,, 
\label{app4}
\ee
where spatial indices are raised and lowered using $\gamma_{AB}$ and
its inverse $\gamma^{AB}$. From their definitions it is clear that
$\vphi$, $\psi$, $\om$ and $\chi$ each correspond to one scalar degree
of freedom, the transverse $3$-vectors $\com_A$ and $\cchi_A$ each
correspond to two functional degrees of freedom, and the transverse
tracefree $3$-tensor $\cchi_{AB}$ corresponds also to two functional
degrees of freedom. This totals to $10$ degrees of freedom, of which
$4$ are coordinate degrees of freedom which can be arbitrarily fixed,
which is what one means by a gauge choice. For example, the
\emph{conformal Newtonian} or \emph{longitudinal} or \emph{Poisson
  gauge} \cite{bruni,mukhanov,note-cNnomen} is defined by the
conditions 
\be
\om = 0 = \chi~~;~~ \cchi^A=0\,.
\label{app5}
\ee
For the metric \eqref{app1} we have at first order,
\be
\sqrt{-{\rm det}\,g} = a^4(\eta)\left(1+ \vphi - 3\psi\right)\,.
\label{app6}
\ee
The inverse of metric \eqref{app1}, correct to first order, has the
components 
\begin{align}
&g^{00} = -\frac{1}{a^2}(1-2\vphi) ~~;~~ g^{0A} = \frac{1}{a^2}\om^A
\,, \nonumber\\
&g^{AB} = \frac{1}{a^2}\left((1+2\psi)\gamma^{AB} - \chi^{AB} \right)\,.
\label{app7}
\end{align}
Denoting $\Cal{H} = (a^\prime/a)$, the prime denoting a derivative
with respect to conformal time $\eta$, the first order accurate
Christoffel symbols are 
\begin{align}
&\G^0_{00} = \Cal{H} + \vphi^\prime ~~;~~ \G^0_{0A} = \p_A\vphi +
  \Cal{H}\om_A\,, \nonumber\\ 
&\nonumber\\
&\G^A_{00} = \p^A\vphi + \om^{A\prime} + \Cal{H}\om^A\,, \nonumber\\ 
&\nonumber\\
&\G^0_{AB} = \left(\Cal{H}-\psi^\prime-2\Cal{H}(\vphi+\psi)
  \right)\gamma_{AB} - \nabla_{(A}\om_{B)} \nonumber\\
&\ph{\G^0_{AB} = (\Cal{H}-)} + \frac{1}{2}\chi^\prime_{AB} +
  \Cal{H}\chi_{AB}\,, \nonumber\\ 
&\nonumber\\
&\G^A_{0B} = \left(\Cal{H} - \psi^\prime\right)\delta{}^A_B +
  \frac{1}{2}\left(\nabla_B\om^A - \nabla^A\om_B \right)
 + \frac{1}{2}\chi^{A\prime}_B\,, \nonumber\\ 
&\nonumber\\
&\G^A_{BC} = {}^{(3)}\bar\G^A_{BC} - \left(\delta{}^A_B\p_C\psi +
  \delta{}^A_C\p_B\psi - \gamma_{BC}\p^A\psi \right) \nonumber\\
&\ph{\G^A_{BC} = } -\Cal{H}\om^A\gamma_{BC} +
  \frac{1}{2}\left(\nabla_C\chi^A_B + \nabla_B\chi^A_C -
  \nabla^A\chi_{BC} \right)\,,  
\label{app8}
\end{align}
where ${}^{(3)}\bar\G^A_{BC}$ denotes the Christoffel connection
associated with the homogeneous $3$-metric $\gamma_{AB}$. 

{\bf Gauge transformations :} While the concept of gauge
transformations can be described in a rather sophisticated language
using pullback operators between manifolds \cite{bruni}, for our
purposes it suffices to implement a gauge transformation using the
simpler notion of an infinitesimal coordinate transformation (also
known as the ``passive'' point of view)
\cite{note6-zalasteady}. Hence, denoting the coordinates and
perturbation functions in the new gauge with a tilde (i.e. 
$\ti{x}^a$, $\ti{\vphi}$, $\ti{\com_A}$, and so on), we have 
\be
\ti{x}^a = x^a + \xi^a(x) ~~;~~ x^a = \ti{x}^a - \xi^a \,, 
\label{app9}
\ee
where the infinitesimal $4$-vector $\xi^a$ can be decomposed as
\be
\xi^a = \left(\xi^0, \xi^A\right) = \left( \alpha, \p^A\beta + d^A
\right)\,, 
\label{app10}
\ee
where $\alpha$ and $\beta$ are scalars and $d^A$ is a transverse
$3$-vector satisfying $\nabla_Ad^A = 0$.

It is then easy to show that if this transformation is assumed to
change the metric \eqref{app1} by changing only the perturbation
functions but leaving the background intact (a so-called ``steady''
coordinate transformation), then the old perturbations and the new are 
related by \cite{bruni}
\begin{align}
&\vphi = \ti{\vphi} + \alpha^\prime + \Cal{H}\alpha \,,\nonumber\\
&\psi = \ti{\psi} - \frac{1}{3}\nabla^2\beta - \Cal{H}\alpha
\,,\nonumber\\ 
&\om = \ti{\om} - \alpha + \beta^\prime \,,\nonumber\\
&\com^A = \ti{\com^A} + d^{A\prime} \,,\nonumber\\
&\chi = \ti{\chi} + 2\beta \,,\nonumber\\
&\cchi^A = \ti{\cchi^A} + d^A \,,\nonumber\\
&\cchi_{AB} = \ti{\cchi_{AB}}\,.
\label{app11}
\end{align}
The last equality shows that the transverse tracefree tensor
perturbations are gauge invariant. They correspond to gravitational
waves. 
\subsection{Time evolution of the background : An iterative approach} 
\label{sec:MG:sub:iter} 
\noindent
Before we move on to deriving formulae for the correlation terms
\eqref{MG8} in terms of perturbation functions in the metric, there is
one issue which merits discussion. The cosmological perturbation
setting, together with the paradigm of averaging, presents us with a
rather peculiar situation. On the one hand, the time evolution of the
scale factor is needed in order to solve the equations satisfied by
the perturbations. Indeed, the standard practice is to fix the time
evolution of the background once and for all, and to use this in
solving for the evolution of the perturbations. On the other hand, the
evolution of the \emph{perturbations} (i.e. -- the inhomogeneities) is
needed to compute the correlation terms appearing in \eqns{MG7}. Until
these terms are known, the behaviour with time of the scale factor 
cannot be determined; and until we know the scale factor as a function
of time, we cannot solve for the perturbations. Note that this is a
generic feature independent of all details of the averaging
procedure. 

It would appear therefore, that we have reached an impasse. To clear
this hurdle, one can try the following iterative approach :
Symbolically denote the background as $a$, the inhomogeneities as
$\vphi$, and the correlation objects as $C$. Note that $a$, $\vphi$
and $C$ all refer to functions of time. We start with a chosen
background, say a standard flat FLRW background with radiation,
baryons and cold dark matter (CDM), and solve for the perturbations in
the usual way, \emph{without} accounting for the correlation terms
$C$. In other words, for this ``zeroth iteration'', we artificially
set $C$ to zero and obtain $a^{(0)}$ and $\vphi^{(0)}$ using the
standard approach (see e.g. \Cite{dodelson}). Clearly, since the
``true'' background (say $a_\ast$) satisfies \eqns{MG7} \emph{with a
  nonzero} $C$, we have in general $a^{(0)}\neq a_\ast$. Now, using
the solution $\vphi^{(0)}$, we can calculate the zeroth iteration
correlation objects $C^{(0)}$ by applying the prescription to be
developed later in this paper. As a first correction to the solution
$a^{(0)}$, we now solve for a new background $a^{(1)}$, with the
\emph{known functions} $C^{(0)}$ acting as sources in \eqns{MG7}. This
first iteration will then yield a solution $\vphi^{(1)}$ for the
inhomogeneities, and hence a new set of correlation terms $C^{(1)}$,
and this procedure can be repeatedly applied. [See however the first
  paper in \Cite{coley} for an alternative approach exclusively using
  averaged quantities in solving the full MG equations.] Pictorially, 
\be
a^{(0)} \longrightarrow \vphi^{(0)} \longrightarrow C^{(0)}
\longrightarrow a^{(1)} \longrightarrow \vphi^{(1)} \longrightarrow
\ldots 
\label{MG10}
\ee
As for convergence, if perturbation theory \emph{is} in fact a good
approximation to the real universe, then one can expect that the
correlation terms will tend to be small compared to other background
objects, and will therefore not affect the background significantly at
each iteration, leading to rapid convergence. On the other hand, if
the correlation terms are large, this procedure may not converge and
one might expect a breakdown of the perturbative picture itself
\cite{note1-tp}. We will see that in the linear regime of
cosmological perturbation theory, the correlation terms do in fact
remain negligibly small. 

\section{The Averaging Operation and Gauge Related Issues}
\label{sec:avgVP}
\noindent
In this section, we will describe the details of the MG (spatial)
averaging procedure adapted to the setting of cosmological PT. 

\subsection{Volume Preserving (VP) Gauges and the Correlation Scalars} 
\label{sec:avgVP-A}
\noindent
It will greatly simplify the discussion if we start with symbolic
calculations which allow us to see the broad structure of the objects
we are after. Since the correlation objects in \eqns{MG7} depend only
on derivatives of the metric, we will primarily deal with metric
fluctuations; matter perturbations will only come into play when 
solving for the actual dynamics of the system. Before dealing with the
issue of which gauge to choose in order to set the condition
\eqref{MG-avgcond}, we will show that irrespective of this choice, the
leading order contribution to the correlations requires knowledge of
only \emph{first order} perturbation functions. 

We will use the following symbolic notation :
\begin{itemize}
\item Inhomogeneous connection: \G 
\item FLRW connection: \GF 
\item Perturbation in the connection : $\dG \equiv \G -
  \GF = \dg + \dG^{(2)}  + \ldots$ \vskip .05in  
\item Coordination bivector : $\W \equiv 1 + \dW = 1 + \dw + \dW^{(2)}
  + \ldots$ 
\item Bilocal extension of the connection : \Gb
\item Inhomogeneous part of the bilocal extension of the connection : $\dGb
  \equiv \Gb - \GF = \dgb + \dGb^{(2)} + \ldots$
\item Correlation object : ${}^{(grav)}T$
\end{itemize}
The integer superscripts denote the order of perturbation. The form of
the coordination bivector arises from the fact that in perturbation
theory, \emph{in the spatial averaging limit}, a transformation from
an arbitrary gauge to a VP one can be achieved by an infinitesimal
coordinate transformation. By a VP gauge we mean a gauge in which the
metric determinant is independent of the \emph{spatial coordinates} to
the relevant order in PT, but may be a function of time. It can be
shown that such a function of time (which will typically be some power
of the scale factor), is completely consistent with all definitions
and requirements of MG in the spatial averaging limit. An easy way of
seeing this is to note that in any averaged quantity, the metric
determinant appears in two integrals, one in the numerator and the
other in the denominator (which gives the normalising volume). In the
``thin time slicing'' approximation we are using to define the
averaging, any overall time dependent factor in the metric determinant
therefore cancels out. Also, a fully volume preserving coordinate
system can clearly be obtained from any VP gauge as defined above, by
a suitable rescaling of the time coordinate. It is not hard to show
that in the thin time slicing approximation, this gives the same
coordination bivector \Wxx{a}{b}\ as the VP gauge definition above.

To see that first order perturbations are sufficient to calculate
${}^{(grav)}T$ to leading order, we only have to note that the
background connection \GF\ satisfies
\be
\avg{\GF} = \GF \,,
\label{avgVP1}
\ee
and that the structure of ${}^{(grav)}T$ is given by
\eqn{MG4}. ${}^{(grav)}T$ then reduces to
\be
{}^{(grav)}T  = \avg{\dGb^2} - \avg{\dGb}^2\,,
\label{avgVP2}
\ee
which is exact. Clearly, the correlation is quadratic in the
perturbation as expected, and hence to leading order, \dGb\ above can
be replaced by \dgb. 

\eqns{avgVP1} and \eqref{avgVP2} treat the averaging operation at a
conceptual level only. To make progress however, we also need to
prescribe how to \emph{practically} impose the averaging assumption 
\be
\avg{\Gb} = \GF ~~\text{i.e.}~~ \avg{\dGb}=0\,,
\label{avgVP3}
\ee
in any given perturbative context. This requires some discussion
since, for example, the bilocal extension of the connection
\Gb\ has the structure 
\be
\Gb = \G\W + \Wi(\p + \pp)\W \,,
\label{avgVP4}
\ee
where \p\ is a derivative at $x$ and \pp\, a derivative at
$x^\prime$. [The reader is referred to \Cite{zala1,zala2} for the
detailed expression. Suffice it to note that this structure ensures
that the averaged connection has the correct transformation
properties.] The actual MG averaging operation in general is therefore
a rather involved procedure. Additionally, it is also necessary to
address certain gauge related issues. 

To clarify the situation, let us start with a fictitious setting in
which the geometry has \emph{exactly} the flat FLRW form, with no
physical perturbations. Clearly, if we work
in the standard comoving coordinates in which the metric $\gamma_{AB}$
of \eqn{FLRW-metric} is simply $\gamma_{AB}=\delta_{AB}$, then since
these coordinates are volume preserving in the sense described above,
the coordination bivector becomes trivial. The averaging involves a
simple integration over $3$-space, and we can easily see that
\eqn{avgVP1} is \emph{explicitly} recovered. 

Now suppose that we perform an infinitesimal coordinate
transformation, \emph{after} imposing \eqn{avgVP1}. Since the
averaging operation is covariant, then from the point of view of a 
general coordinate transformation, \emph{both} sides of \eqn{avgVP1}
will be affected in the same way. However, suppose that we had
performed the transformation \emph{before} imposing \eqn{avgVP1}.
In the language of cosmological PT, we would then be dealing with some
``pure gauge'' perturbations around the fixed, spatially homogeneous
background. If we did not know that these perturbations were pure
gauge, we might naively construct the nontrivial coordination bivector
for this metric, compute the bilocal extension of the connection
according to \eqn{avgVP4} and try to impose \eqn{avgVP3}. This would
be incorrect since these perturbations were arbitrarily generated and
need not average to zero (for example they could be positive definite
functions). In order to maintain consistency, it is then necessary to
ensure in practice that the averaging condition \eqref{avgVP3} is
applied only to gauge invariant fluctuations (which is rather obvious
in hindsight).   

There is another problem associated with the structure of the
coordination bivector, even when there \emph{are} real, gauge
invariant inhomogeneities present. Note from \eqn{MG-W} that the
coordination bivector has the structure 
\be
W = \left .\frac{\p x}{\p x_V}\right|_{x^\prime} \left .\frac{\p
  x_V}{\p x}\right|_x 
\label{avgVP5}
\ee
where $x$ denotes the coordinates we are working in and $x_V$ a set of
VPCs. In perturbation theory (in the spatial averaging limit) we will
have, at leading order,
\be
x = x_V - \xi~~;~~ x_V = x + \xi\,,
\label{avgVP6}
\ee
where $\xi$ symbolically denotes an infinitesimal $4$-vector defining
the transformation, and hence
\be
(\p x_V)/(\p x) = 1 + \p\xi
\label{avgVP7}
\ee
and so on, which gives us
\be
\W = 1 - (\p\xi)|_{x^\prime} + (\p\xi)|_x + \ldots = 1 + \dw + \ldots 
\label{avgVP8}
\ee
Now when we compute a quantity such as \avg{\GF\dw}\, which appears in
the expression \eqref{avgVP4} for \avg{\Gb}\,, we will be left with a
fluctuating ($\vec{x}$-dependent) term of the form $\GF(\avg{\p\xi} -
\p\xi)$, where $\vec{x}$ denotes the $3$ spatial coordinates. Hence if
we try to impose \eqn{avgVP3} we will ultimately be left with
equations of the type  
\be
\avg{f}(\vec{x}) - f(\vec{x}) = 0
\label{avgVP9}
\ee
for some functions derived from the inhomogeneities which we have
collectively denoted $f$. In other words, consistency would seem to
demand that the inhomogeneities vanish in this coordinate system,
which is of course not desirable. 

It therefore appears that we are forced to impose \eqn{avgVP3}
\emph{in a volume preserving gauge}, since by definition, only in such
a gauge will we have $\W=1$ exactly. We emphasize that this is a
purely practical aspect related to defining the averaging operation,
and is completely decoupled from, e.g. the choice of gauge made when
studying the time evolution of perturbations. We are in no way
breaking the usual notion of gauge invariance by choosing an averaging
operator. The conditions \eqn{avgVP9} now reduce to the form  
\be
\avg{f_{VPC}}(\vec{x}) = 0
\label{avgVP10}
\ee
which are far more natural than \eqn{avgVP9}. The averaging condition
is now unambiguous, \emph{but depends on a choice of the VP gauge
  which defines the averaging operation},
an issue we shall discuss in the next subsection. For now, all we can
assert is that this VP gauge must be such that in the \emph{absence}
of gauge invariant fluctuations, it must reduce to the standard
comoving (volume preserving) coordinates of the background geometry as
in \eqn{FLRW-metric}. This of course is simply the statement that the
VP gauge must be \emph{well defined} and must not contain any residual
degrees of freedom.

The averaging operation now takes on an almost trivial form -- to
leading order it is easy to show that for any quantity
$f(\eta,\vec{x})$ (with or without indices), the average of $f$ in a
VP gauge in the spatial averaging limit, is given by
\be
\avg{f}(\eta,\vec{x}) = \frac{1}{V_L}\int_{\Cal{V}(\vec{x})}{ d^3y
  f(\eta,y)} \,, 
\label{avgVP11}
\ee
where the integral is over a spatial domain $\Cal{V}(\vec{x})$ with a 
constant volume $V_L$. The spatial coordinates are the comoving
coordinates of the background metric, and at leading order the
boundaries of $\Cal{V}(\vec{x})$ can be specified in a straightforward
manner as, e.g., 
\be
\Cal{V}(\vec{x}) = \{\vec{y}~|~x^A-L/2 < y^A < x^A+L/2, A=1,2,3. \}\,,
\label{avgVP12}
\ee
where $L$ is a comoving scale over which the averaging is performed
(in which case $V_L=L^3$). The averaging definition can be written
more compactly in terms of a window function $W_L(\vec{x},\vec{y})$ as  
\begin{align}
&\avg{f}(\eta,\vec{x}) = \int{d^3y W_L(\vec{x},\vec{y})f(\eta,\vec{y})}
\,, \nonumber\\
& \int{d^3y W_L(\vec{x},\vec{y})} = 1\,,
\label{avgVP13}
\end{align}
where $W_L(\vec{x},\vec{y})$ vanishes everywhere except in the region 
$\Cal{V}(\vec{x})$, with the integrals now being over all space. This
expression will come in handy when working in Fourier space, as we
shall do in later sections. 

A couple of comments are in order at this stage. Firstly, we have not
specified the magnitude of the averaging scale $L$. The general
philosophy is that this scale must be large enough that a single
averaging domain encompasses several realisations of the random
inhomogeneous fluctuations, and small enough that the observable
universe contains a large number of averaging domains. However, as we
will show later in \Sec{sec:corrscal}, if one is ultimately interested
in quantities which are formally averaged over an ensemble of
realisations of the universe (as is usually done in interpreting
observations), then the actual value of the averaging scale becomes
irrelevant. 

This brings us to the second issue. The above discussion is valid only
in the situation where there are no fluctuations at arbitrarily large
length scales, since in the presence of such fluctuations the
averaging condition \eqref{avgVP3} loses meaning (in such a
situation it would be impossible to isolate the background from the
perturbation by an averaging operation on any finite length
scale). Indeed, we shall see a manifestation of this restriction in
\Sec{sec:corrscal}, where the correlation scalars will be seen to
diverge in the presence of a nonzero amplitude at arbitrarily large
scales, of the power spectrum of metric fluctuations. 

We will end this subsection by explicitly writing out the averaging
condition in an ``unfixed VP'' gauge, to be defined below, and also
writing the correlation terms appearing in \eqn{MG7}, in this gauge. 
As we can see from \eqn{app6}, the basic condition to be satisfied by 
a VP gauge is
\be
\ti{\vphi} = 3\ti{\psi}\,.
\label{avgVP14}
\ee
Hereafter, all VP gauge quantities will be denoted using a
tilde. This should not be confused with the similar notation that was
used for the bilocal extension in \Sec{sec:MG:sub:recap}, which will
not be needed in the rest of the paper. $\ti{\vphi}$ and $\ti{\psi}$
are the scalar potentials appearing in the perturbed FLRW metric
\eqref{app1}. The single condition \eqref{avgVP14} leaves $3$ degrees
of freedom to be fixed, in order to completely specify the VP gauge
one is working with. The MG formalism by itself does not prescribe a
method to choose a particular VPC system; in fact this freedom of
choice of VPCs is an inherent part of the formalism. We shall return to
this issue in the next subsection. For now we define the ``unfixed VP 
(uVP) gauge'' by the single requirement \eqref{avgVP14}, with $3$
unfixed degrees of freedom, and present the expressions for the
averaging condition and the correlation scalars, with this choice. 

It is straightforward to determine the consequences of requiring
\eqn{MG-avgcond} to hold, with the right hand side corresponding to
the FLRW connection in conformal coordinates, and remembering that the
coordination bivector (in the spatial averaging limit) is now trivial
(see \eqn{MG-W-VP}). Together with some additional reasonable
requirements, namely
\be
\avg{\nabla^2s} = 0 = \avg{\nabla^2\p_As}\,,
\label{avgVP15}
\ee
for any scalar $s(\eta,\vec{x})$, the averaging condition in the
uVP gauge reduces to
\begin{align}
&\avg{\ti{\psi}} = 0 ~~;~~ \avg{\p_A\ti{\psi}} = 0 =
  \avg{\ti{\psi}^\prime}\,, \nonumber\\ 
&\avg{\ti{\om}_A} = 0 = \avg{\ti{\om}^\prime_A} ~~;~~
  \avg{\ti{\chi}^\prime_{AB}} = 0\,, \nonumber\\ 
&\avg{\nabla_C\ti{\chi}^A_B} + \avg{\nabla_B\ti{\chi}^A_C} -
  \avg{\nabla^A\ti{\chi}_{BC}} = 0\,, \nonumber\\ 
&\avg{\nabla_A\ti{\om}_B} = \avg{\nabla_B\ti{\om}_A} =
  \Cal{H}\avg{\ti{\chi}_{AB}} \,,
\label{avgVP-avgcond}
\end{align}
where we have used the expressions in \eqn{app8} with the uVP
condition \eqref{avgVP14}. We will also make the additional reasonable
requirement that 
\be
\avg{\ti{\chi}_{AB}} = 0\,,
\label{avgVP16}
\ee
using which it is easy to see that the perturbed FLRW metric
\eqref{app1} and its inverse \eqref{app7}, in the uVP 
gauge, \emph{both} on averaging reduce to their respective homogeneous
counterparts, namely
\be
\avg{g_{ab}} = g_{ab}^{(FLRW)} ~~;~~ \avg{g^{ab}} = g^{ab}_{(FLRW)}\,.  
\label{avgVP-metricavg}
\ee
Using these results, the expressions \eqref{MG8} simplify to give, in
the uVP gauge,
\begin{subequations}
\begin{align}
&\nonumber\\
&\Cal{P}^{(1)} = \frac{1}{a^2} \left[ 6\avg{(\ti{\psi}^\prime)^2} +
    \avg{\nabla_{[A}\ti{\om}_{B]} \nabla^{[A}\ti{\om}^{B]}}
    \right. \nonumber\\
&\ph{\Cal{P}^{(1)} = \frac{1}{a^2} [ 6\avg{(\ti{\psi}^\prime)^2}]
    +}      \left. - \frac{1}{4}\avg{\ti{\chi}_{AB}^\prime
      \ti{\chi}^{AB\prime}} \right] \,, \label{avgVP-corrscal-a}\\ 
&\nonumber\\
&\Cal{S}^{(1)} = \frac{1}{a^2} \left[ -10\avg{\p_A\ti{\psi}
      \p^A\ti{\psi}} - 2\avg{\p_A\ti{\psi}\nabla_B\ti{\chi}^{AB}}
    \right. \nonumber\\
&\ph{\Cal{S}^{(1)}} \left. + \frac{1}{4}
    \avg{\nabla^B\ti{\chi}^{AC}\left(2\nabla_A\ti{\chi}_{BC}
    - \nabla_B\ti{\chi}_{AC}\right) }  \right]
  \,,\label{avgVP-corrscal-b}\\ 
&\nonumber\\
&\Cal{P}^{(1)} + \Cal{P}^{(2)} = \frac{1}{a^2} \left[
    6\avg{(\ti{\psi}^\prime)^2} -
    24\Cal{H}\avg{\ti{\psi}^\prime\ti{\psi}} \right. \nonumber\\
&\ph{\Cal{P}^{(1)} + \Cal{P}^{(2)}} \left . -
    \avg{\ti{\psi}^\prime\nabla^2\ti{\om}} +
    \frac{1}{2}\avg{\ti{\chi}^\prime_{AB}\nabla^A\ti{\om}^B}
    \right. \nonumber\\ 
&\ph{\Cal{P}^{(1)} + \Cal{P}^{(2)}} \left. - \frac{1}{4}
    \avg{\ti{\chi}^\prime_{AB}\left(\ti{\chi}^{AB\prime} +
      2\Cal{H}\ti{\chi}^{AB} \right)
    }\right]\,, \label{avgVP-corrscal-c}\\ 
&\nonumber\\
&\Cal{S}^{(2)} = \frac{1}{a^2} \left[
    3\avg{\ti{\om}^{A\prime}\p_A\ti{\psi}} +
    \Cal{H}\avg{\ti{\om}^A\ti{\om}^\prime_A} \right]
  \,, \label{avgVP-corrscal-d} 
\end{align}
\label{avgVP-corrscal}
\end{subequations}
where square brackets denote antisymmetrization.

\subsection{Choice of VP Gauge} 
\label{sec:ginvar}
\noindent
In this subsection we will prescribe a choice for the VP gauge which 
defines the averaging operation. In general, the class of volume
preserving coordinate systems for any spacetime, is very large (see
\Cite{mars} for a detailed characterisation). We have so far managed
to pare it down by requiring that the VP gauge we choose should reduce
to the standard FLRW coordinates in the absence of fluctuations. It
turns out to be somewhat difficult to go beyond this step, since there
does not appear to be any unambiguously clear guiding principle
governing this choice. We will therefore motivate a choice for the VP
gauge based on certain details of cosmological PT which one knows from
the standard treatments of the subject.

In particular, we shall make use of certain nice properties of the
conformal Newtonian or longitudinal or Poisson
gauge, which is defined by the conditions \eqref{app5}
\cite{bruni,mukhanov} (henceforth we shall refer to this gauge as the
cN gauge for short). Since this gauge is well defined and has no
residual degrees of freedom, all the nonzero perturbation functions in
the cN gauge, namely $\vphi$, $\psi$, $\com_A$ and $\cchi_{AB}$ in the
notation of \Sec{app:pertavg}, are equal to gauge invariant
objects. This is trivially true for $\cchi_{AB}$, as seen in the last
equation in \eqref{app11}. For the rest, note that in any arbitrary
unfixed gauge, the following combinations are gauge invariant at first
order 
\begin{align}
&\Phi_B = \vphi + \frac{1}{a}\p_\eta\left[a\left( \om -
  \frac{1}{2}\chi^\prime \right)\right] \,,\nonumber\\
&\Psi_B = \psi - \Cal{H}\left(\om - \frac{1}{2}\chi^\prime \right) +
\frac{1}{6}\nabla^2\chi\,,\nonumber\\
&{\hat V}_A = \com_A - \cchi^\prime_A\,,
\label{ginvar2}
\end{align}
which can be easily checked using \eqns{app11}, and in the cN gauge,
$\om$, $\chi$ and $\cchi_A$ all vanish. Here $\Phi_B$ and $\Psi_B$ are
the Bardeen potentials \cite{bardeen} (upto a sign), and $\Psi_B$ in
particular has the physical interpretation of giving the gauge
invariant \emph{curvature perturbation}, which is the quantity on
which initial conditions are imposed post inflation
\cite{liddle-lyth}.  

Additionally, it is also known that the cN gauge \emph{for the metric}
remains stable even during structure formation, when matter
inhomogeneities have become completely nonlinear. (See \Cite{wald} for
an intuitive description of why this is so, and \Cite{SphColl} for an
explicit demonstration in a toy model of structure formation.) We
believe that this is a strong argument in favour of using the cN gauge
to define a VP gauge which will then define the averaging
operation in the perturbative context. This will ensure that this
``truncated'' averaging operation, defined for first order PT, will
remain valid \emph{at leading order} even during the nonlinear epochs
of structure formation. 

To implement this in practice, consider a transformation from the cN
gauge to the uVP gauge defined by \eqn{avgVP14}. The transformation
equations \eqref{app11} reduce to  
\begin{align}
&\alpha^\prime + 4\Cal{H}\alpha + \nabla^2\beta = \vphi - 3\psi
  \,,\nonumber\\ 
&\ti{\psi} = \frac{1}{3}\vphi - \alpha^\prime - \Cal{H}\alpha
  \,,\nonumber\\ 
&\ti{\om} = \alpha - \beta^\prime \,,\nonumber\\
&\ti{\com^A} = \com^A - d^{A\prime} \,,\nonumber\\
&\ti{\chi} = - 2\beta \,,\nonumber\\
&\ti{\cchi^A} = - d^A \,,\nonumber\\
&\ti{\cchi_{AB}} = \cchi_{AB}\,.
\label{ginvar3}
\end{align}
Recall that to completely specify a VP gauge, we need to fix $3$
degrees of freedom in the uVP gauge. Our requirement regarding the
``well defined''-ness of the VP gauge, forces us to set $d^A=0$, and
to choose $\alpha$ and $\beta$ such that they vanish in the case where
$\vphi=0=\psi$. 

This has fixed $2$ degrees of freedom, in addition to the condition
\eqref{avgVP14} which is just the definition of the uVP gauge, and has
hence not yielded a uniquely specified VP gauge. To do this, we shall
make the following additional requirement. Since we are dealing with a
spatial averaging, it seems reasonable to require that the VP gauge
being used to define the averaging, should be ``as close as possible''
to the cN gauge in terms of \emph{time slicing}, and for this reason
we shall set the function $\alpha$ to zero.  

To summarize, the VP gauge chosen is defined in terms of the gauge
transformation functions $\xi^a = (\alpha,\p^A\beta+d^A)$ between the
cN gauge and the VP gauge, by the following relations
\be
\alpha = 0 = d^A \,,
\label{ginvar-VP1}
\ee
and
\begin{subequations}
\begin{align}
&\ti{\vphi} = 3\ti{\psi} = \vphi\,,\label{ginvar-VP2-a}\\ 
&\nabla^2\beta = \vphi - 3\psi\,,\label{ginvar-VP2-b}\\
&\ti{\om} = -\beta^\prime ~~;~~ \ti{\chi} = -2\beta
  \,,\label{ginvar-VP2-c}\\ 
&\ti{\cchi}_A = 0\,,\label{ginvar-VP2-d}\\
&\ti{\com}_A = \com_A ~~;~~ \ti{\cchi}_{AB} =
  \cchi_{AB}\,,\label{ginvar-VP2-e} 
\end{align}
\label{ginvar-VP2}
\end{subequations}
where the function $\beta$ is restricted not to contain any nontrivial
solution of the homogeneous (Laplace) equation $\nabla^2\beta=0$. 

Having made this choice for the VP gauge, we are now assured that all
averaged quantities which we compute are gauge invariant : our choice
ensures that the averaging procedure does not introduce any pure gauge
modes, and the philosophy of ``steady'' coordinate transformations
ensures that all background objects are, by assumption, unaffected by
gauge transformations. In particular, the correlation objects in
\eqns{MG8} are all gauge invariant. This is
different from the gauge invariance conditions derived in the first
paper of \Cite{avg-bran}, where the background was also taken to
change under gauge transformations at second order in the
perturbations. It is at present not clear how these results are
related to ours. 

Note that all these arguments are valid at first order in PT, which is
sufficient for our present purposes. A consistent treatment at second
order would require more work, although as long as one is interested
only in the leading order effect, these arguments are expected to go
through. 

\section{The Correlation Scalars}
\label{sec:corrscal}
\noindent
With the VP gauge choice defined by \eqns{ginvar-VP2}, it is
straightforward to rewrite the correlation objects in
\eqns{avgVP-corrscal} (which are in the uVP gauge) in terms 
of the perturbation functions in the cN gauge. We will restrict the
subsequent calculations in this paper to the case where there are no
transverse vector perturbations, i.e.,
\be
\com_A = 0\,,
\label{corrscal1}
\ee
in the cN gauge. This is a reasonable choice since such vector
perturbations, even if they are excited in the initial conditions,
decay rapidly and do not source the other perturbations at first order
\cite{dodelson}. 

In addition, for simplicity (and to keep this paper concise) we will
choose to ignore the gauge invariant tensor perturbations as well,
\be
\cchi_{AB} = 0\,.
\label{corrscal2}
\ee
It will be an interesting excercise to account for the effects
of gravitational waves in the correlation scalars, however we will
leave this to future work. Thus, the results presented here apply only
to \emph{scalar perturbations}. 

In terms of the scalar perturbations in the cN gauge, for a flat FLRW
background, the correlation objects \eqref{avgVP-corrscal} reduce to
\begin{subequations}
\begin{align}
&\Cal{P}^{(1)} = \frac{1}{a^2} \bigg[ \, 2\avg{(\psi^\prime)^2} + 
    \avg{\left(\vphi^\prime - \psi^\prime\right)^2} \nonumber\\
&\ph{\Cal{P}^{(1)} = \frac{1}{a^2} [ 6\avg{}]+}    -
    \avg{\left(\nabla_A\nabla_B\beta^\prime\right)
      \left(\nabla^A\nabla^B\beta^\prime\right)} \,  \bigg]
  \,, \label{corrscal-a}\\   
&\Cal{S}^{(1)} = -\frac{1}{a^2} \bigg[ 6\avg{\p_A\psi\p^A\psi} +
    \avg{\p_A(\vphi-\psi)\p^A(\vphi-\psi)}  \nonumber\\
&\ph{\Cal{S}^{(1)}= -\frac{1}{a^2}}  -
    \avg{(\nabla_A\nabla_B\nabla_C\beta)
      (\nabla^A\nabla^B\nabla^C\beta)}  \bigg]  
  \,,\label{corrscal-b}\\ 
&\nonumber\\
&\Cal{P}^{(1)} + \Cal{P}^{(2)} = \frac{1}{a^2} \bigg[
    \avg{\vphi^\prime(\vphi^\prime-\psi^\prime)}
     \nonumber\\    
     &\ph{\Cal{P}^{(1)} + }- 2\Cal{H}\left\{\,
    \avg{\vphi^\prime\vphi} -  \avg{\psi^\prime\psi}  +
    \avg{\psi^\prime(\vphi-\psi)}  \right.  \nonumber\\   
&\ph{\Cal{P}^{(1)} + } \left.  +  
    \avg{\psi(\vphi^\prime-\psi^\prime)}  +
    \avg{(\nabla_A\nabla_B\beta)(\nabla^A\nabla^B\beta^\prime)}\, 
    \right\}  \bigg]\,, \label{corrscal-c}\\  
&\Cal{S}^{(2)} = -\frac{1}{a^2} \bigg[
    \avg{\p^A\beta^{\prime\prime} \left(\p_A\vphi -
      \Cal{H}\p_A\beta^\prime \right)}  \bigg] 
  \,, \label{corrscal-d} 
\end{align}
\label{corrscal}
\end{subequations}
where $\beta$ is defined in \eqn{ginvar-VP2-b}.

Since we are working with a flat FLRW background, it becomes
convenient to transform our expressions in terms of Fourier space
variables. This will also highlight the problem with large scale
fluctuations which was mentioned in \Sec{sec:avgVP}. We will use the
following Fourier transform conventions : For any scalar function
$f(\eta,\vec{x})$, its Fourier transform $f_{\vec{k}}(\eta)$ satisfies 
\begin{align}
f(\eta,\vec{x}) &= \int{\frac{d^3k}{(2\pi)^3}e^{i\vec{k}\cdot\vec{x}}
  f_{\vec{k}}(\eta)} \,,\nonumber\\
f_{\vec{k}}(\eta) &= \int{d^3x
  e^{-i\vec{k}\cdot\vec{x}}f(\eta,\vec{x})} \,.
\label{corrscal3}
\end{align}
Consider an average of a generic quadratic product of two scalars
$f^{(1)}(\vec{x})$ and $f^{(2)}(\vec{x})$ where we have suppressed the
time dependence since it simply goes along for a ride. Using the
definition \eqref{avgVP13}, and keeping in mind that the scalars are
real, it is easy to show that we have  
\be
\avg{f^{(1)}f^{(2)}}(\vec{x}) =
\int{\frac{d^3k_1d^3k_2}{(2\pi)^6}W^\ast_L(\vec{k}_1-\vec{k}_2,\vec{x})
f^{(1)}_{\vec{k}_1}f^{(2)\ast}_{\vec{k}_2}} \,,
\label{corrscal4}
\ee
where $W_L(\vec{k},\vec{x})$ is the Fourier transform of the window
function $W_L(\vec{x},\vec{y})$ on the variable $\vec{y}$, and the
asterisk denotes a complex conjugate.

In the present context, the functions $f^{(1)}$ and $f^{(2)}$ will
typically be derived in terms of the initial random fluctuations in
the metric $\vphi_{\vec{k}i}$ which are assumed to be drawn from a
\emph{statistically homogeneous and isotropic} Gaussian distribution
with some given power spectrum. In order to ultimately make contact
with observations, it seems necessary to 
perform a formal ensemble average over all possible realisations of
this initial distribution of fluctuations. The statistical homogeneity
and isotropy of the initial distribution implies that the functions
$f^{(1)}$ and $f^{(2)}$ will satisfy a relation of the type
\be
\eavg{f^{(1)}_{\vec{k}_1}f^{(2)\ast}_{\vec{k}_2}} =
(2\pi)^3\delta^{(3)}(\vec{k}_1 - \vec{k}_2)P_{f_1f_2}(|\vec{k}_1|)\,,
\label{corrscal5}
\ee
for some function $P_{f_1f_2}(k,\eta)$ which is derivable in terms of the
initial power spectrum of metric fluctuations, and where $\eavg{...}$
denotes an ensemble average and $\delta^{(3)}(\vec{k})$ is the Dirac
delta distribution. 

Applying an ensemble average to \eqn{corrscal4} introduces a Dirac
delta which forces $\vec{k}_1=\vec{k}_2$. Further, the normalisation
condition on the window function in \eqn{avgVP13} implies that we have
\be
W_L(\vec{k}=0,\vec{x}) = 1\,,
\label{corrscal6}
\ee
which means that all dependence on the averaging scale and domain
drops out, and we are left with
\be
\eavg{\avg{f^{(1)}f^{(2)}}} = \int{\frac{d^3k}{(2\pi)^3}
  P_{f_1f_2}(k)}\,.  
\label{corrscal7}
\ee
Note however, that the right hand side of \eqn{corrscal7} is precisely
what we would have obtained, had we treated the spatial average
$\avg{...}$ to be the ensemble average $\eavg{...}$ to begin
with. Therefore for all practical purposes, we are justified in
replacing all the spatial averages in the expressions for the
correlation scalars \eqref{corrscal}, by ensemble averages. 

It is convenient to define the transfer function $\Phi_k(\eta)$ via
the relation  
\be
\vphi_{\vec{k}}(\eta) = \vphi_{\vec{k}i}\Phi_k(\eta)\,.
\label{corrscal8}
\ee
For the calculations in this paper, we shall
assume that the cN gauge scalars $\vphi(\eta,\vec{x})$ and
$\psi(\eta,\vec{x})$ are equal 
\be
\vphi(\eta,\vec{x}) = \psi(\eta,\vec{x})\,,
\label{corrscal9}
\ee
a choice which is valid in first order PT when anisotropic stresses
are negligible (see \Cite{dodelson}). This simplifies many of the
expressions we are dealing with. The Fourier transform of $\beta$ can
be written, using \eqns{ginvar-VP2-b} and \eqref{corrscal9}, as
\be
\beta_{\vec{k}}(\eta) = \frac{2}{k^2}\vphi_{\vec{k}}(\eta)\,.
\label{corrscal10}
\ee
Finally, in terms of the transfer function $\Phi_k(\eta)$ and
the initial power spectrum of metric fluctuations defined by
\be
\eavg{\vphi_{\vec{k_1}i}\vphi^\ast_{\vec{k_2} i}} = (2\pi)^3
\delta^{(3)}(\vec{k_1} - \vec{k_2}) P_{\vphi i}(k_1)\,,
\label{corrscal-powspec}
\ee
the correlation scalars \eqref{corrscal} can be written as (compare
Eqns (58)-(61) of \Cite{behrend})
\begin{subequations}
\begin{align}
&\nonumber\\
&\Cal{P}^{(1)} = -\frac{2}{a^2} \int{\frac{dk}{2\pi^2}
    k^2P_{\vphi i}(k) \left(\Phi_k^\prime \right)^2 } 
\,, \label{corrscal-kspace-a}\\   
&\nonumber\\
&\Cal{S}^{(1)} = -\frac{2}{a^2} \int{\frac{dk}{2\pi^2}
  k^2P_{\vphi i}(k) \left( k^2 \Phi_k^2\right)}
  \,,\label{corrscal-kspace-b}\\ 
&\nonumber\\
&\Cal{P}^{(1)} + \Cal{P}^{(2)} = -\frac{8\Cal{H}}{a^2}
  \int{\frac{dk}{2\pi^2} k^2P_{\vphi i}(k) \left( \Phi_k\Phi_k^\prime 
    \right) } 
\,, \label{corrscal-kspace-c}\\  
&\nonumber\\
&\Cal{S}^{(2)} = -\frac{2}{a^2}   \int{\frac{dk}{2\pi^2}
  k^2P_{\vphi i}(k) \Phi_k^{\prime\prime}\left( \Phi_k -
  \frac{2\Cal{H}}{k^2}\Phi_k^\prime \right) }  
  \,. \label{corrscal-kspace-d} 
\end{align}
\label{corrscal-kspace}
\end{subequations}
These expressions highlight the problem of having a finite amplitude
for fluctuations at arbitrarily large length scales ($k\to0$), which
was mentioned in \Sec{sec:avgVP}. As a concrete example, consider the
frequently discussed Harrison-Zel'dovich scale invariant spectrum
\cite{scalinvarpowspec} which satisfies the condition
\be
k^3P_{\vphi i}(k) = {\rm constant}\,.
\label{scale-invar}
\ee
\eqns{corrscal-kspace} now show that if the transfer function
$\Phi_k(\eta)$ has a finite time derivative at large scales (as it
does in the standard scenarios -- see the next section), then the
correlation objects $\Cal{P}^{(1)}$, $\Cal{P}^{(2)}$ and
$\Cal{S}^{(2)}$ all diverge due to contributions from the $k\to0$
regime. This demonstrates the importance of having an initial power
spectrum in which the amplitude dies down sufficiently rapidly on
large length scales (which is a known issue, see
\Cite{liddle-lyth}). Perturbation theory cannot adequately describe
the behaviour of inhomogeneities with arbitrarily large length scales
\cite{note2-zala}. Keeping this in mind, we shall concentrate on
initial power spectra which display a long wavelength cutoff
\cite{note3-behrend}. Models of inflation leading to such power
spectra have been discussed in the literature \cite{cutoff-theory},
and more encouragingly, analyses of WMAP data seem to indicate
that such a cutoff in the initial power spectrum is in fact realised
in the universe \cite{cutoff-obsvns}.  

A final comment before proceeding to explicit calculations : In
addition to picking up nontrivial correlation corrections in the
cosmological equations, the averaging formalism also requires that the
``cross-correlation'' constraints in \eqns{MG9} be satisfied. In the
absence of vector and tensor modes, it is straightforward to show that 
the statistical homogeneity and isotropy of the scalar metric
fluctuations implies that these constraints are identically
satisfied. It will be an interesting excercise to analyse the
conditions imposed by these constraints in the presence of tensor
modes; we leave this for future work. 
\section{Worked out examples}
\label{sec:exs}
\noindent
We will now turn to some explicit calculations of the correlation
integrals (backreaction), which will show that the magnitude of
the effect remains negligibly small for most of the evolution duration
in which linear PT is valid. At early times, linear PT is valid at
practically all scales including the smallest scales at which we wish
to apply General Relativity. As matter fluctuations grow, the small
length scales progressively approach nonlinearity, and linear PT
breaks down at these scales. As we will see, however, by the time a
particular length scale becomes nonlinear, its contribution to the
amplitude of the \emph{metric} fluctuations correspondingly becomes 
negligible. In practice therefore, one can extend the linear
calculation well into the matter dominated era, with the expectation
that the order of magnitude of the various integrals will not change 
significantly due to nonlinear effects (see also the discussion in the
last section). 

The model we will use is the standard Cold Dark Matter (sCDM) model
consisting of radiation and CDM \cite{dodelson}. We will neglect
the contribution of baryons, and at the end we shall discuss the
effects this may have on the final results. We shall also discuss,
without explicit calculation, the effects which the introduction of a
cosmological constant is likely to have. In the following,
$\Omega_r$ and $\Omega_m$ denote the density parameters of
radiation and CDM respectively at the present epoch $\tau_0$, with
$\tau$ denoting cosmic time. $\Omega_r$ is assumed to contain
contributions from photons and $3$ species of massless,
out-of-equilibrium neutrinos. At the ``zeroth iteration'' (see
\Sec{sec:MG:sub:iter}) we have  
\be
\left(\frac{1}{a}\frac{da}{d\tau}\right)^2 = H^2(a) = H_0^2\left[ 
  \frac{\Omega_m}{a^3} + \frac{\Omega_r}{a^4} \right]  \,,
\label{exs1}
\ee
where $H_0$ is the standard Hubble constant, the scale factor is
normalised so that $a(\tau_0)=1$, and $\Cal{H}$ and $H$ are related by  
\be
\Cal{H}(a) = aH(a)\,.
\label{exs2}
\ee
The comoving wavenumber corresponding to the scale which enters at the
matter radiation equality epoch, is given by
\be
k_{eq} = a_{eq}H(a_{eq}) =
H_0\left(\frac{2\Omega^2_m}{\Omega_r}\right)^{1/2} \sim H_0\cdot
10^{5/2}\,, 
\label{exs3}
\ee
where we have set (see \Cites{kolb-turner,dodelson} for details)
\begin{align}
\Omega_r &= \Omega_{photon} + 3\Omega_{neutrino}
\nonumber\\ 
&= \Omega_{photon}\left(1 +
3\cdot\frac{7}{8}\left(\frac{4}{11}\right)^{4/3} \right) \nonumber\\
&= 4.15\times10^{-5}h^{-2}\,, 
\label{exs4}
\end{align}
where $h$ is the dimensionless Hubble parameter defined by $H_0 =
100h$ km/s/Mpc. For all calculations we shall set $h=0.72$
\cite{hst-key}. 
\subsection{EdS background and non-evolving potentials}
\label{sec:exs:sub:eds}
\noindent
Before dealing with the full model (which requires a numerical
evolution) let us consider the simpler situation, described by an
Einstein-deSitter (EdS) background, with negligible radiation and a
nonevolving potential $\vphi = \vphi(\vec{x})$ (which is a consistent
solution of the Einstein equations in the sCDM model at least at
subhorizon scales at late times \cite{dodelson}). Although not fully
accurate, this example requires some very simple integrals and will
help to give us a feel for the structure and magnitude of the
backreaction. 

With a constant potential, the only correlation object which survives
is $\Cal{S}^{(1)}$, which evolves like $\sim a^{-2}$, where the scale
factor refers to the ``zeroth iteration''. The constant of
proportionality can be written in terms of the BBKS transfer function
$T_{BBKS}(k/k_{eq})$ \cite{bbks,dodelson}, to give 
\be
\Cal{S}^{(1)} = -\frac{2}{a^2}\int{\frac{dk}{2\pi^2}k^4
  P_{\vphi i}(k)T^2_{BBKS}(k/k_{eq})}\,,
\label{exs-eds1}
\ee
where we have \cite{bbks}
\begin{align}
T_{BBKS}(x) &= \frac{\ln{[1+0.171x]}}{(0.171x)} \bigg[
  1 + 0.284x  \nonumber\\
& + (1.18x)^2 + (0.399x)^3 + (0.490x)^4 \bigg]^{-0.25}\,,
\label{exs-eds2}
\end{align}
where $x\equiv (k/k_{eq})$.

The integral in \eqn{exs-eds1} is well-behaved even in the presence of 
power at arbitrarily large scales, for a (nearly) scale invariant
spectrum. Since we are only looking for an estimate, we shall evaluate
the integral in the absence of a large scale cutoff, and leave a more
accurate calculation for the next subsection. For the initial spectrum
given by 
\be
\frac{k^3P_{\vphi i}(k)}{2\pi^2} = A (k/H_0)^{n_s - 1}\,,
\label{exs-eds3}
\ee
where the scalar spectral index $n_s$ is close to unity, the integral
in \eqn{exs-eds1} can be easily performed numerically and has the
order of magnitude  
\be
\int{\frac{dk}{2\pi^2}k^4 P_{\vphi i}(k)T^2_{BBKS}(k/k_{eq})} \sim
A\left(k_{eq}\right)^2 \sim AH_0^2\cdot10^5 \,, 
\label{exs-eds4}
\ee
upto a numerical prefactor of order $1$. Since the amplitude
of the power spectrum is $A\sim 10^{-9}$ \cite{liddle-normalisn}, the 
overall contribution of the backreaction is  
\be
\frac{\Cal{S}^{(1)}}{H_0^2} \sim -\frac{1}{a^2}(10^{-4})\,.
\label{exs-eds5}
\ee
Now, as long as the correlation objects give a negligible
backreaction to the usual background quantities, when we proceed with
the \emph{next} iteration, the effect of the backreaction on the
evolution of the \emph{perturbations} will also remain negligible (at
least at the leading order). Hence in practice there will be
essentially no difference between the zeroth iteration and first
iteration perturbation functions. This amounts to saying that when the
backreaction is negligible, convergence to the ``true'' solution for
the scale factor \emph{at the leading order}, is essentially achieved
in a single calculation. 

\subsection{Radiation and CDM without baryons} 
\label{sec:exs:sub:nobar}
\noindent
Let us now turn to the full sCDM model (without baryons). An
analytical discussion of this model in various regions of
$(k,\eta)$-space, can be found e.g. in \Cite{dodelson}. Since we are
interested in integrals over $k$ across a range of epochs $\eta$, it
is most convenient to solve this model numerically. It is further
convenient to use $(\ln a)$ in place of $\eta$, as the variable with
which to advance the solution. Also, it is useful to introduce
transfer functions like $\Phi_k(\eta)$ for all the relevant
perturbation functions in exactly the same manner (see
\eqn{corrscal8}), namely by pulling out a factor of
$\vphi_{\vec{k}i}$, since the initial conditions are completely
specified by the initial metric perturbation. For a generic
perturbation function $s_{\vec{k}}(\eta)$ (other than the metric 
fluctuation $\vphi_{\vec{k}}$) the transfer function
corresponding to $s$ will be denoted by a caret, so that 
\be
s_{\vec{k}}(\eta) = \vphi_{\vec{k}i}\hat s_k(\eta)
\label{exs-nobar1}
\ee
The relevant Einstein equations can be brought to the following closed
set of first order ordinary differential equations (adapted from
Eqns. (7.11)-(7.15) of \Cite{dodelson}),  
\begin{subequations}
\begin{align}
&\frac{\p\Phi_k}{\p(\ln a)}= -\left[ \left(1 +
    \frac{K^2}{3E^2}\right)\Phi_k \right. \nonumber\\
&\ph{\frac{\p\Phi_k}{\p(\ln a)}= -1} \left. +
    \frac{1}{2E^2a}\left(\Omega_m\cdelta_k + 
    \frac{4}{a}\Omega_r\ctheta_{0k}  \right)
    \right]\,, \label{exs-nobar2a}\\ 
&\nonumber\\
&\frac{\p\cdelta_k}{\p(\ln a)} = -\frac{K}{E}\cv_k +
  3\frac{\p\Phi_k}{\p(\ln a)}  \,,\label{exs-nobar2b}\\
&\nonumber\\
&\frac{\p\ctheta_{0k}}{\p(\ln a)} = -\frac{K}{E}\ctheta_{1k} +
  \frac{\p\Phi_k}{\p(\ln a)}  \,,\label{exs-nobar2c}\\ 
&\nonumber\\
&\frac{\p\ctheta_{1k}}{\p(\ln a)} = \frac{K}{3E}\left(\ctheta_{0k} + 
  \Phi_k \right)   \,,\label{exs-nobar2d}\\ 
&\nonumber\\
&\frac{\p \cv_k}{\p(\ln a)}= -\cv_k + \frac{K}{E}\Phi_k
  \,.\label{exs-nobar2e} 
\end{align}\label{exs-nobar2}
\end{subequations}
Here we have introduced the dimensionless variables
\be
K\equiv \frac{k}{H_0} ~~;~~ E(a) \equiv \frac{\Cal{H}(a)}{\Cal{H}_0} =
\frac{\Cal{H}(a)}{H_0} \,,
\label{exs-nobar3}
\ee
and the various perturbation functions are defined as follows :
$\delta_k$ is the $k$-space density contrast of CDM, $\Theta_{0k}$ and
$\Theta_{1k}$ are the monopole and dipole moments respectively of the
$k$-space temperature fluctuation of radiation, and $(-iV_k)$ is the
$k$-space peculiar velocity scalar potential of CDM (i.e., the real
space peculiar velocity is $v_A = \p_Av$ where $v$ is the Fourier
transform of $(-iV_k)$). 

Assuming adiabatic perturbations, the initial conditions satisfied by
the transfer functions at $a=a_i$ are (adapted from Ch. 6 of
\Cite{dodelson}) 
\begin{align}
&\Phi_k(a_i) = 1 ~~;~~ \cdelta_k(a_i) = -\frac{3}{2} ~~;~~
\ctheta_{0k}(a_i) = -\frac{1}{2} \,, \nonumber\\ 
&\cv_k(a_i) = 3\ctheta_{1k}(a_i) = \frac{1}{2}\frac{K}{E(a_i)}\,.
\label{exs-nobar-init}
\end{align}
We choose $a_i = 10^{-16}$, which corresponds to an initial background
radiation temperature of $T\sim10^3$GeV. While this is not as far back in
the past as the energy scale of inflation (which is closer to
$\sim10^{15}$Gev), it is on the edge of the energy scale where known
physics begins \cite{liddle-lyth}. This makes \eqn{exs4} unrealistic
since we have ignored all of Big Bang Nucleosynthesis and also the fact
that neutrinos were in equilibrium with other species at temperatures
higher than about $1$Mev. However the modifications due to these
additional details are not expected to drastically change the final
results, and these assumptions lead to some simplifications in the
code used. The goal here is only to demonstrate an application of the
formalism; more realistic calculations accounting for the effects of
baryons can also be performed (see, e.g. Behrend et
al. \cite{behrend} who incorporate these effects for the
post-recombination era, albeit in the Buchert formalism).  
\begin{figure}[t]
\includegraphics[width=0.48\textwidth]{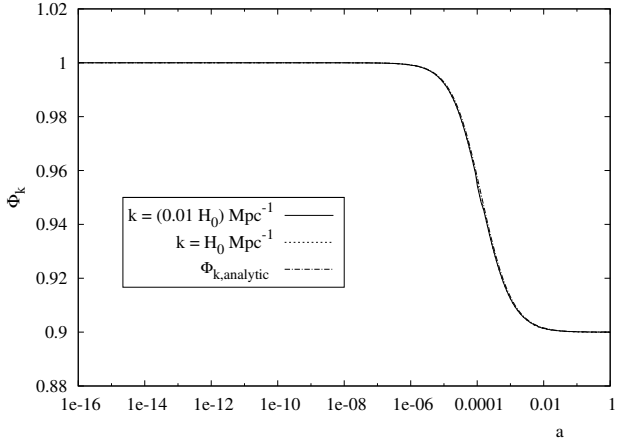}
\caption{\small The evolution of two large scale modes. Also shown is
  the Kodama-Sasaki analytical solution in the large scale limit
  $k\eta\ll1$, given by \eqn{exs-numres1}.}  
\label{fig1}
\end{figure}

In order to partially account for the fact that inflationary initial
conditions are actually set much earlier than $a=10^{-16}$, we impose
an absolute \emph{small wavelength} cutoff at the scale which enters
the horizon at the initial epoch which \emph{we} have chosen. In the
above notation this corresponds to setting
$K_{max}=E(a_i)\sim10^{13}$. This makes sense since scales satisfying
$K\gg K_{max}$ have already entered the horizon and decayed
considerably by the epoch $a=10^{-16}$. There is a source of error due
to ignoring scales $K\gtrsim K_{max}$ which have not yet decayed
significantly, but this error rapidly decreases with time as
progressively larger length scales enter the horizon and decay. [In
  fact, in practice to compute the integrals at any given epoch
  $a=a_\ast$, one only needs to have followed the evolution of modes
  with $K<\,\sim5000E(a_\ast)$ : more on this in the next subsection.]
More important is the cutoff at \emph{long} wavelengths, which we set
at $K_{min} = 1$ (corresponding to $k_{min}=H_0$), which is firstly a
natural choice given that $H_0^{-1}$ is the only large scale in the
system, and is secondly also guided by analyses of CMB data which
have detected such a cutoff \cite{cutoff-obsvns}. We will see that
reducing $K_{min}$ even by a few orders of magnitude, does not
affect the final qualitative results significantly. 

\subsubsection{Numerical Results}  
\label{sec:exs:sub:nobar:sub:numres}
\noindent
Equations \eqref{exs-nobar2} with initial conditions
\eqref{exs-nobar-init} were solved with a standard $4$th order
Runge-Kutta integrator with adaptive stepsize control (based on the
algorithm given in \Cite{NumRec}). For the integrals in
\eqns{corrscal-kspace}, only the function $\Phi_k(a)$ needs to be
tracked accurately. Hence, although $\ctheta_{0k}$ and $\ctheta_{1k}$
are difficult to follow accurately beyond the matter radiation
equality $a_{eq} = (\Omega_r/\Omega_m) \simeq 8\times10^{-5}$ due to
rapid oscillations, the integrals can still be reliably computed since
$\ctheta_{0k}$ and $\ctheta_{1k}$ do not significantly affect the
evolution of $\Phi_k$ in the matter dominated era (as seen in
\eqn{exs-nobar2a}). 
\begin{figure}[t]
\centering
\includegraphics[width=0.48\textwidth]{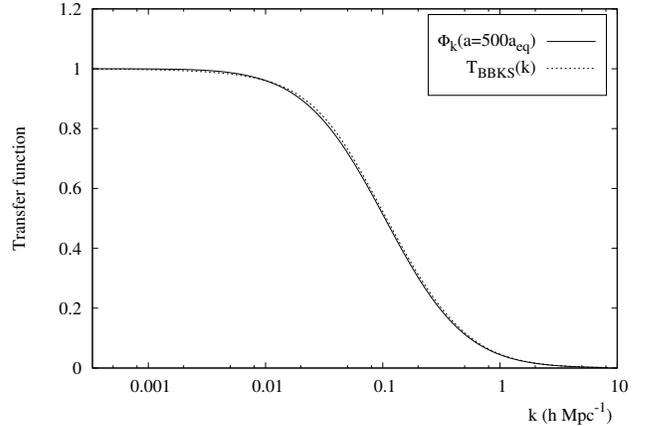}
\caption{\small The transfer function $\Phi_k$ normalised by its
  constant value at large scales, at the epoch $a=500a_{eq}$. The
  dotted line is the BBKS transfer function \eqref{exs-eds2}.} 
\label{fig2}
\end{figure}

To see that known results are being reproduced by the code, consider
\figs{fig1} and \ref{fig2} as examples. \fig{fig1} shows the
evolution of two scales corresponding to $K=1$ ($k=H_0\,{\rm
  Mpc}^{-1}$) and $K=0.01$. The first enters the horizon at the
present epoch, while the second remains superhorizon for the entire
evolution, satisfying $k\eta\ll1$. In this limit an analytical
solution exists in the sCDM model, due to Kodama and Sasaki
\cite{kodama,dodelson}, given by 
\be
\Phi_k(y) = \frac{1}{10y^3}
\left[16\sqrt{1+y} + 9y^3 + 2y^2 -8y -16 \right] \,,
\label{exs-numres1}
\ee
where $y\equiv a/a_{eq}$, and this function is also shown. Clearly all
the curves in \fig{fig1} are practically identical.

\fig{fig2} shows the function $\Phi_k$ normalised by its (constant)
value at large scales, at the epoch $a=500a_{eq}\simeq 0.04$,
(which is well into the matter dominated era). The dotted line is the
BBKS fitting form given in \eqn{exs-eds2} with $k_{eq}$ given by
\eqn{exs3}. 

To numerically estimate the integrals in \eqns{corrscal-kspace}, the
values of $\Phi_k$ and its first and second derivatives with respect
to $(\ln a)$ are needed across a range of $K$ values. For reference,
note that the following relations hold for a generic function of time
$w(\eta)$, 
\be
\frac{dw}{d\eta} = a\Cal{H}\frac{dw}{da} = \Cal{H}\frac{dw}{d(\ln
  a)} \,.
\label{exs-numres2}
\ee
Based on the earlier discussion, the initial power spectrum
$P_{\vphi i}(k)$ is taken to satisfy
\be
\frac{k^3P_{\vphi i}(k)}{2\pi^2} = A ~,~~ {\rm for\,\,} H_0 < k < 
k_{max}=\Cal{H}(a_i) \,,
\label{exs-numres3}
\ee
and zero otherwise, and we set
\be
A = 1.0\times10^{-9}\,,
\label{exs-numres4}
\ee
which, for the sCDM model follows from the convention (see Eqn.(6.100)
of \Cite{dodelson}) $A = (5\delta_H/3)^2$ with
$\delta_H\approx2\times10^{-5}$ (see, e.g. \Cite{liddle-normalisn}).  
\begin{figure}[t]
\includegraphics[width=0.48\textwidth]{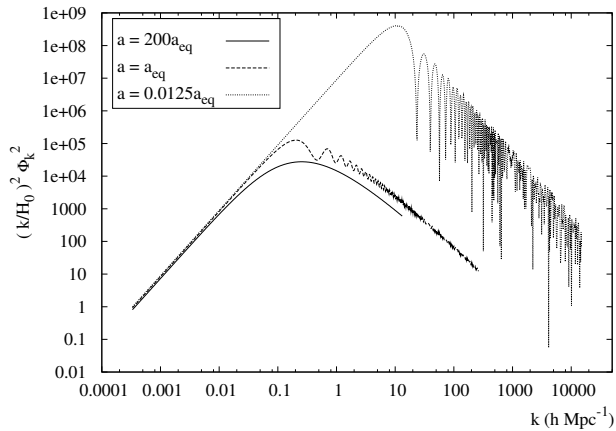}
\caption{\small The dimensionless integrand of $\Cal{S}^{(1)}$, namely
the function $(k/H_0)^2\Phi_k^2$, at three sample values of the scale
factor. The function dies down rapidly for large $k$, with the value
at some $k$ being progressively smaller with increasing scale
factor. The declining behaviour of the curves for $a=a_{eq}$ and
$a=200a_{eq}$ extrapolates to large $k$.}  
\label{fig3}
\end{figure}

\begin{figure}[t]
\includegraphics[width=0.48\textwidth]{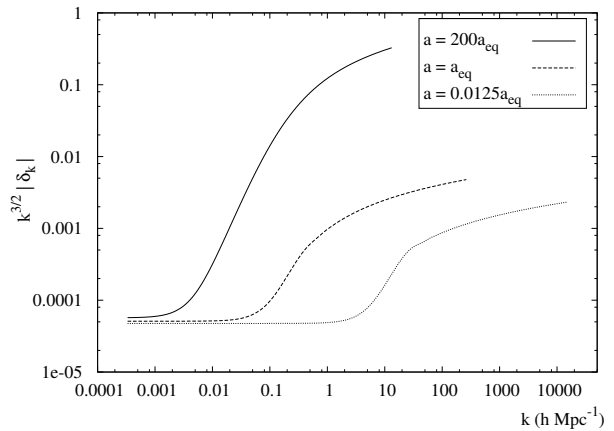}
\caption{\small The dimensionless CDM density contrast. Together with
  \fig{fig3} this shows that nonlinear scales do not impact the
  backreaction integrals significantly. }   
\label{fig4}
\end{figure}

Consider \figs{fig3} and \ref{fig4}, which highlight two issues
discussed earlier. \fig{fig3} shows the integrand of $\Cal{S}^{(1)}$
at three sample epochs, and we see that the integrand dies down
rapidly at increasingly smaller $k$ values for progressively later
epochs. (The other integrands, not displayed here, also show this
rapid decline for large $k$.) [We have not shown the integrand at the
  later two epochs for all values of $k$ since this was
  computationally expensive, but the declining trend of the curves 
  can be extrapolated to large $k$, which is well understood
  analytically \cite{dodelson}.] This justifies the statement in
the beginning of this section, that at any epoch $a_\ast$ it is
sufficient to have followed the evolution of scales satisfying 
$K<5000E(a_\ast)$ for computing the integrals. Secondly, \fig{fig4}
shows the behaviour of $k^{3/2}|\delta_k| = A^{1/2}|\cdelta_k|$ at the
same three epochs, and comparing with \fig{fig3} we see that at any
epoch, the region of $k$-space where linear PT has broken down, does
not contribute significantly to the integrals. This is in line with
the conjecture in \Cite{paddy} that the effects of the backreaction
should remain small since the mass contained in the nonlinear scales
is subdominant.

Due to the structure of the integrals and the chosen initial power
spectrum, it is convenient to compute the integrands in
\eqns{corrscal-kspace} equally spaced in $(\ln K)$, and then perform
the integrals using the extended Simpson's rule \cite{NumRec}. If
$2^N+1$ points are used to evaluate a given integral, resulting in a
value $\Cal{I}_N$ say, then the error can be estimated by computing
the integral with $2^{N-1}+1$ points to get $\Cal{I}_{N-1}$, and
estimating the relative error as $|\Cal{I}_{N-1}/\Cal{I}_N| -1$. With
$N=10$, the estimated errors in all the integrals at all epochs were
typically less than $0.1\%$. A bigger error is incurred in computing
the integrand itself at any given epoch, leading to estimated errors
of order $\sim1\%$ in $\Cal{S}^{(1)}$, $\Cal{P}^{(1)}$ and
$\Cal{P}^{(1)}+\Cal{P}^{(2)}$, with a larger error in $\Cal{S}^{(2)}$
as explained below.

The second derivative $\p^2\Phi_k/\p(\ln a)^2$ proves to be difficult
to track numerically. At early times, when most scales are
superhorizon, the Kodama-Sasaki analytical solution
\eqref{exs-numres1} is a good approximation for most values of
$k$. Using this one can see that at early times the value of the
derivative is numerically very small, and is difficult to reliably
estimate due to roundoff errors. For this reason the integral
$\Cal{S}^{(2)}$ could not be accurately estimated at early
times. However, the structure of the integrand of $\Cal{S}^{(2)}$
\eqref{corrscal-kspace-d} shows that the largest contribution comes
from large (superhorizon) scales (the small scales being subdominant
due to the presence of $\Phi_k$ and $1/k^2$). An analysis using the
Kodama-Sasaki solution then shows in a fairly straightforward manner
that the behaviour of the backreaction term is
$|\Cal{S}^{(2)}/H^2|\sim 10^{-6}(a/a_{eq})(H_0/k_{min})^2$ for our
choices of parameters, where $(a/a_{eq})\ll1$. At intermediate times
around $a\sim a_{eq}$ and later, although it becomes computationally
expensive to obtain convergent values for the second derivative at all
relevant scales \cite{note4-accuracy}, moderately good accuracy
($1$-$5\%$) can be achieved.   
\begin{figure}[t]
\includegraphics[width=0.48\textwidth]{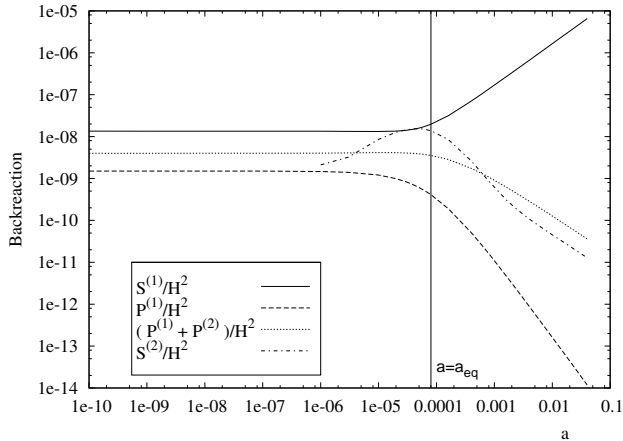}
\caption{\small The correlation scalars (``backreaction'') for the
  sCDM model, normalised by $H^2(a)$. $\Cal{S}^{(1)}$, $\Cal{P}^{(1)}$
  and $\Cal{S}^{(2)}$ are negative definite and their magnitudes have
  been plotted. The vertical line marks the epoch of matter radiation
  equality $a=a_{eq}$.}    
\label{fig5}
\end{figure}

The results are shown in \fig{fig5}, in which the
magnitudes of the correlation integrals of \eqn{corrscal-kspace},
normalised by the Hubble parameter squared $H^2(a) = (\Cal{H}/a)^2$
are plotted as a function of the scale factor in a log-log plot. The
values for $\Cal{S}^{(2)}$ are shown only for epochs later than
$a\simeq0.01a_{eq}\sim10^{-6}$. We see that at
all epochs, the correlation terms remain negligible compared to the
chosen zeroth iteration background. Also, in the radiation dominated
epoch all the correlation scalars (except $\Cal{S}^{(2)}$ whose
evolution couldn't be accurately obtained) track the $\sim a^{-4}$
behaviour of the background radiation density (see also the second
paper in \Cite{avg-bran}). The discussion above shows however that the
magnitude of $\Cal{S}^{(2)}$ is far smaller than the other
backreaction functions at early times, for a cutoff at
$k_{min}=H_0$. On the other hand, in the matter dominated epoch
$\Cal{S}^{(1)}$ dominates the backreaction and settles into a
curvature-like $\sim a^{-2}$ behaviour (note that in the matter  
dominated epoch we have $H^2\sim a^{-3}$). This can also be compared
with the results of \Cite{siegel}. As for the signs of the
correlations, $\Cal{S}^{(1)}$, $\Cal{S}^{(2)}$, and $\Cal{P}^{(1)}$
are negative throughout the evolution while
$\Cal{P}^{(1)}+\Cal{P}^{(2)}$ is positive throughout.   

Finally, a few comments regarding the effects of ignoring baryons,
nonlinear corrections, etc. Including baryons in the problem (with a
background density parameter of $\Omega_b\simeq0.05$) will lead to a
significant suppression of small scale power (by introducing pressure
terms which will tend to wipe out inhomogeneities) and also a small
suppression of large scale power. This effect causes a
(downward) change in the late time transfer function of roughly
$15$-$20\%$ \cite{dodelson}, and therefore cannot increase the
contribution of the backreaction. Quasi-linear corrections can lead to
significant changes in the transfer function, but do not cause shifts
by several orders of magnitude (see \Cite{halo} and references
therein). Hence accounting for changes due to quasi-linear behaviour
will also not increase the magnitude of the backreaction by a large
amount (see also \Cite{behrend}).  As for effects from fully nonlinear
scales, we have seen that these can be expected to remain small, or at
least not orders of magnitude larger than those from linear scales
(see also the discussion in the last section, and \Cite{siegel}).  

Adding a cosmological constant (and retaining a flat background
geometry) will change the qualitative features of the correlation
functions by shifting the scale $k_{eq}$ (due to a reduced $\Omega_m$,
which will also increase the power spectrum amplitude \cite{dodelson},
but again not by orders of magnitude). Also, the late time behaviour
of the correlation scalars will be affected since the potential
$\Phi_k$ will decay at late times instead of remaining
constant. Regardless, the backreaction is expected to remain small
even in this case (which is also indicated by the calculations of
Behrend et al. \cite{behrend} in the Buchert framework
\cite{buchert}). 
 
\section{Discussion}
\label{sec:discuss}
\noindent
This paper has presented an analysis of cosmological perturbation
theory (PT) in the fully covariant averaging framework of
Zalaletdinov's Macroscopic Gravity (MG) \cite{zala1,zala2} and its
restriction to spatial averaging \cite{spatavglim}. While this
framework is generally covariant, the issue of gauge dependence in
perturbation theory introduces certain subtleties in the problem. We
have shown that, provided one takes seriously the idea that the
cosmological background must be defined by an averaging procedure
\cite{zala-pert}, it is possible to attach a gauge invariant meaning
to the averaging condition and the corresponding correlation objects
which appear as corrections to the cosmological equations. While there
remains considerable freedom in an explicit choice of the averaging
operator (through a choice of the volume preserving gauge used in its
definition), this freedom can be fixed by some additional requirements
based on knowledge of cosmological PT in the standard
implementations. In particular we have seen that properties of the
conformal Newtonian or Poisson gauge can be used to motivate a fully
specified choice of the averaging operation adapted to first order
PT. 

One prerequisite to the formulation of a consistent averaging
framework in the context of perturbation theory, is the absence of
perturbative fluctuations with arbitrarily large wavelengths, since
such fluctuations would render meaningless the notion of recovering a 
homogeneous background on averaging. This problem also manifested
itself in the correlation integrals \eqref{corrscal-kspace}, which
diverge in the presence of a finite amplitude of fluctuations as the
wavenumber $k\to0$. Accordingly, all the calculations of this paper
have assumed that the \emph{initial} power spectrum of metric
fluctuations has a sharp cutoff at the scale corresponding to $k=H_0$,
a hypothesis which is in fact supported by analysis of CMB data
\cite{cutoff-obsvns}. 

The main purpose of this paper was to lay down the formalism of MG in
a language most convenient from the point of view of cosmological
PT. This was accomplished by writing \eqns{avgVP-corrscal},
\eqref{corrscal} and the Fourier space version
\eqref{corrscal-kspace} (with certain simplifying assumptions 
regarding vector and tensor perturbations which can if needed be
relaxed in a completely straightforward manner). This was supplemented
by calculations in the sCDM model \cite{dodelson} (which is the flat
FLRW model with radiation and Cold Dark Matter but no Dark Energy)
with the additional simplification of ignoring the baryons. The
analytical results of \Sec{sec:exs:sub:eds} as well as the more
detailed numerical results of \Sec{sec:exs:sub:nobar} show that the
correlation objects or ``backreaction'' remain negligibly small upto
epochs corresponding to a scale factor of $a\sim0.01$. While the
calculations ignored corrections from quasi-linear and nonlinear
scales, these are not expected to contribute dramatically to the
correlations obtained here, an expectation which is justified by the
work of Behrend et al. \cite{behrend} and further by the calculation
in \Cite{NLcoll-MG} (see below). 

We have seen that by using the framework of MG, we have completely
bypassed the problem mentioned in the Introduction, which one faces
when applying the Buchert framework to cosmological PT, namely of
having to deal with \emph{two} scale factors. Here, one has a single
well-defined scale factor associated with the background metric, and
its evolution can be obtained in an iterative fashion as described in
\Sec{sec:MG:sub:iter}. In practice, we saw that since the backreaction
is small, convergence can be achieved by essentially a single
calculation, at least in the context of first order PT. 

This brings us to a final, and very important issue : What is the
magnitude and behaviour of the backreaction in the fully nonlinear
regime of structure formation? There are (at least) two possible
avenues to approach this question. The first is to set up the problem
in a manner which is suitable for $N$-body simulations. The iterative
approach suggested earlier can presumably be adapted to full-fledged
$N$-body codes as well. While this is a possibility worth pursuing, 
there is also a less involved (but correspondingly less realistic) way
of determining the effect of nonlinearities on the backreaction, which
is to study toy models of structure formation. As mentioned earlier,
such a toy model of spherical collapse was recently presented in
\Cite{SphColl}, and it was shown by an explicit coordinate
transformation, that the metric can be brought to the Newtonian form
\eqref{eq1}. Now, it is worth noting that while all the calculations
of this paper assumed that first order PT is valid, the actual
expressions for the correlation integrals in \eqns{corrscal}
\emph{only} assume that the potentials $\vphi$ and $\psi$ satisfy
$|\vphi|,|\psi|\ll1$. The \emph{dynamics} governing the potentials is
irrelevant at this stage. This means that, as long as one is
interested in leading order effects only,  the expressions in
\eqns{corrscal} \emph{can be directly applied} to any model of
structure formation where the \emph{metric} can be brought to the
conformal Newtonian form \cite{note5-vectortensor}: in particular they
can be applied to the model of \Cite{SphColl}. This has been done in
\Cite{NLcoll-MG}, and one finds that even in the fully nonlinear
regime, the effect of the backreaction remains negligible. In this
context see also \Cite{siegel} for a third approach.

To conclude, all our calculations and arguments seem to indicate that
the averaging of perturbative inhomogeneities in a consistent manner
appears to lead only to very small effects. This does not, however,
mean that the effects do not exist. It remains to be seen whether any
observable consequences of the backreaction may be detectable by
future experiments. \\

\acknowledgments
\noindent
It is a pleasure to thank Prof. T. P. Singh for his patient guidance
and for many insightful conversations, and Prof. T. Padmanabhan for
his encouragement and support, and for fruitful discussions. I am
grateful to Prof. Roustam Zalaletdinov for his patience during an
earlier correspondence which laid the foundations of this work. I
thank Jasjeet Bagla and the astrophysics group at HRI, Allahabad for
their hospitality during a visit, during which this work was
begun. Finally, I am grateful to Alok Maharana, Subhabrata Majumdar
and Rakesh Tibrewala for useful discussions.


\begin{thebibliography}{10}
\bibitem{note-cNnomen}
Note that strictly speaking, ``conformal Newtonian'' refers to the
first order PT gauge whereas ``Poisson'' refers to its higher order
generalisation. We will not make this distinction here.
\bibitem{note4-accuracy}
Convergence was tested by varying a global parameter which dynamically
controls the stepsize during evolution (by stepsize doubling/halving,
see \Cite{NumRec}). The integrals other than $\Cal{S}^{(2)}$ show
convergence at $3$ or more significant digits for all epochs,
whereas convergence can be obtained for $\Cal{S}^{(2)}$ only at epochs
sufficiently close to matter radiation equality, and there only for
$1$-$2$ significant digits, by setting stringent conditions on
stepsize doubling.
\bibitem{note5-vectortensor}
This is true when the model does not have vector and tensor
perturbations. In the presence of such modes one should appropriately
rewrite the expressions in \eqns{avgVP-corrscal}.
\end{thebibliography}
\end{document}